\documentclass[copyright,creativecommons]{eptcs}  
\providecommand{\event}{Infinity 2013}

\usepackage{algorithm2e}
\usepackage{amsmath}
\usepackage{tikz}
\usepackage{amssymb}
\usepackage{etoolbox}
\usepackage{subfigure}
\usepackage{multirow}     
\usepackage{amsthm}

\usetikzlibrary{arrows}
\usetikzlibrary{shapes}
\usetikzlibrary{automata}

\newcommand{\od}{\bar{d}}
\newcommand{\Post}{\mathit{Post}}
\newcommand{\bbfN}{\mathbb{N}}
\newcommand{\until}{\mathbin{U}}
\newcommand{\defn}{\overset{\triangle}{=}}
\newcommand{\ug}{\underline{3}}

\newcommand{\hide}[1]{}
\newtheorem{definition}{Definition}
\newtheorem{example}{Example}
\newtheorem{theorem}{Theorem}
\newtheorem{proposition}{Proposition}
\newtheorem{lemma}{Lemma}

\ifdef{\longversion}{
\newcommand{\whenlongversion}[1]{#1}
}{
\newcommand{\whenlongversion}[1]{}
}

\newcommand{\valueof}[1]{[\![{#1}]\!]}

\title{A Finite Exact Representation of Register Automata Configurations}
\author{Yu-Fang Chen \institute{Academia Sinica, Taiwan} \and Bow-Yaw Wang \institute{Academia Sinica, Taiwan}\and Di-De Yen \institute{Academia Sinica, Taiwan}
}
\def\titlerunning{A Finite Exact Representation of Register Automata Configurations}
\def\authorrunning{Y.-F. Chen, D.-D. Yen \& B.-Y. Wang}

\begin{document}
\maketitle

\begin{abstract}

A register automaton is a finite automaton with finitely
many registers ranging from an infinite alphabet.
Since the valuations of registers are infinite, there are infinitely many configurations.
We describe a technique to classify
infinite register automata configurations into finitely many \emph{exact} representative
configurations. Using the finitary representation, we give an algorithm
solving the reachability problem for register automata. We moreover
define a computation tree logic for register automata and solve
its model checking problem.


\end{abstract}

%
%

\section{Introduction} 


Register automata are generalizations of finite automata to process
strings over infinite alphabets~\cite{R4}. In addition to a finite
set of states, a register automaton has finitely many registers
ranging from an infinite alphabet. When a register automaton reads a
data symbol with parameters from the infinite alphabet, it
compares values of registers and parameters and finite constants, updates registers, and
moves to a new location. Since register automata allow infinitely many
values in registers and parameters, they have been used to model
systems with unbounded data. For instance, a formalization of user
registration and account management in the XMPP protocol is given
in~\cite{R1,R2}. Since user identifiers are not fixed 
\emph{a priori}, models in register automata are more realistic for
the protocol.


Analyzing register automata nonetheless is not apparent. Since there
are infinitely many valuations of registers, the number of
configurations for a register automaton is inherently infinite. 
Moreover, register automata can update a register with values of
registers or parameters in data symbols. The special feature makes
register automata more similar to programs than to classical
automata. Infinite configurations and register updates increase the
expressive power of register automata. They also complicate analysis
of the formalism as well.


In this paper, we develop a finitary representation for configurations
of register automata. As observed in~\cite{R5}, register automata
recognize strings modulo automorphisms on the infinite alphabet. That
is, a string is accepted by a register automaton if and only if the
image of the string under a one-to-one and onto mapping on the infinite
alphabet is accepted by the same automaton. Subsequently, two
valuations of registers are indistinguishable by register automata if
one is the image of the other under an automorphism on the infinite
alphabet. We therefore identify indistinguishable valuations and
classify valuations into finitely many representative
valuations. Naturally, our finitary representation enables
effective analysis on register automata.


The first application of representative valuations is reachability
analysis. Based on representative valuations, we define representative
configurations. Instead of checking whether a given configuration is
reachable in a register automaton, it suffices to check whether
its representative configuration belongs to the finite set of reachable
representative configurations. We give an algorithm to compute
successors of an arbitrary representative configuration. The set of
reachable representative configurations is obtained by fixed point
computation. 


Our second application is model checking on register automata. We
define a computation tree logic (CTL) for register
automata. Configurations in a representative configuration are shown
to be indistinguishable in our variant of computation tree logic. The
CTL model checking problem for register automata thus is solved by the
standard algorithm with slight modifications.


As an illustration, we model an algorithm for the Byzantine generals
problem under an interesting scenario. In the scenario, two loyal
generals are trying to reach a consensus at the presence of a treacherous
general. They would like to know how many soldiers should be sent to
the front line. Since the total number of soldiers is
unbounded,\footnote{This is certainly an ideal simplification. The
number of soldiers of course is bounded by the population of the
empire.} we use natural numbers as the infinite alphabet and
model the algorithm in a register automaton. By the CTL model
checking algorithm, we compute the initial configurations leading to
a consensus eventually.


Our formulation of register automata follows those in~\cite{R1,R2}.
It is easy to show that the expressive power of register automata with constant symbols is no difference from those versions without constants.
A canonical representation theorem similar to Myhill-Nerode theorem for
deterministic register automata is developed in~\cite{R1}. 
In~\cite{R2}, a learning algorithm for register automata is
proposed.
Finite-memory automata is another generalization of finite automata to infinite
alphabets~\cite{R5}.
Finite-memory automata and register automata have the same expressive power.
In~\cite{R5}, we know that the emptiness problem for finite-memory automata is decidable.
Therefore, the reachability problem for register automata is also decidable.
In~\cite{2009:Demri}, it has been shown that the emptiness for register automata is in \emph{PSPACE}. This is done by reducing an emptiness checking problem for register automata to an emptiness problem of a finite transition system over the so called ``abstract states'', which is very similar to the ``equivalence classes'' defined in this paper. However, in their reduction, they did not provide any algorithm to move from one abstract state to another abstract state, which is in fact non-trivial.
In contrast, we provide an algorithm in Section~\ref{section:reachability}.
It has been shown in~\cite{2010:Figueira} that register automata together with a total order over the \emph{alphabet} are equivalent to timed automata.
In fact, the register automata model defined in this paper can be easily extended to support  arbitrary order among alphabet symbols (the order can be partial) and hence is more general then the one defined in~\cite{2010:Figueira}. This because the finite representation of configurations defined in this paper can be extended to describe any finite relations between alphabet symbols by adding more possible values in the matrix. That is, instead of just $0$ and $1$ used in the current paper, we can add more possible values such as $\leq, <, >, \ldots$ to describe a richer relation between alphabet symbols.
A survey on expressive power of various finite
automata with infinite alphabets is given in~\cite{R4}. We model the
algorithm for the Byzantine generals problem~\cite{LSP:82:BGP} presented
in~\cite{W:12:ASBGP}.

The paper is organized as follows. We briefly review register automata
in Section~\ref{section:preliminaries}. 
Section~\ref{section:representative-configurations} presents an exact
finitary representation for configurations. It is followed by the
reachability algorithm for register automata
(Section~\ref{section:reachability}). A computation tree logic for
register automata and its model checking algorithm are given in
Section~\ref{section:model-checking}. We discuss the Byzantine
generals problem as an example
(Section~\ref{section:example}). Finally, we conclude the presentation
in Section~\ref{section:conclusion}.

%
%

\section{Preliminaries}
\label{section:preliminaries}


Let $S$, $S'$, and $S''$ be sets. An \emph{automorphism} on $S$ is
a one-to-one and onto mapping from $S$ to $S$. Given a subset $T$ of $S$, an automorphism $\sigma$ on $S$ is invariant on $T$ if $\sigma (x) = x$ for every $x \in T$. If $f$ is an onto mapping from $S$ to $S'$ and $h$ is a mappings from $S'$ to $S''$, $(h \circ f)$ is a mapping from $S$ to $S''$ that $(h \circ f) (a) = h (f (a))$ for
$a \in S$. We write $S_{n \times n}$ for the set of
square matrices of size $n$ with entries in $S$. 


Let $\Sigma$ be an infinite \emph{alphabet}. A set of constants, denoted by $C$, is a finite subset of $\Sigma$. Let $A$ be a finite set of
\emph{actions}. Each action has a finite \emph{arity}.
A \emph{data symbol} $\alpha (\od_n)$ consists of an action
$\alpha \in A$ and $\od_n = d_1d_2\cdots d_n\in
\Sigma^n$ when $\alpha$ is of arity $n$. A \emph{string} is a sequence
of data symbols.  


Fix a finite set $X$ of \emph{registers}. Define $X' = \{ x' | x \in X
\}$. A \emph{valuation} $v$ is
a mapping from $X$ to $\Sigma$. Since $X$ is finite, we represent a
valuation by a string of $\Sigma^{|X|}$. We write $V_{(X,\Sigma)}$
for the set of valuations from $X$ to $\Sigma$.


Let $P = \{ p_1, p_2, \ldots \}$ be an infinite set of \emph{formal
parameters} and $P_n = \{ p_1, p_2, \ldots, p_n \} \subseteq P$. 
A \emph{parameter valuation} $v_{\od_n}$ is a mapping
from $P_n$ to $\Sigma$ such that $v_{\od_n}(p_i) = d_i$ for every $1
\leq i \leq n$. We write $V_{(P,\Sigma)}$ for the set of parameter
valuations. Obviously, each finite sequence $\od_n \in \Sigma^n$
corresponds to a parameter valuation $v_{\od_n} \in V_{(P, \Sigma)}$.

Given a valuation $v$, a parameter valuation $v_{\od_n}$, and 
$e \in X \cup P_n \cup C$, define
\begin{equation*}
  \begin{array}{rcl}
    \valueof{e}_{v, v_{\od_n}} & = &
    \left\{
      \begin{array}{ll}
        v (e) & \textmd{ if } e \in X\\
        v_{\od_n} (e) & \textmd{ if } e \in P_n\\
        e & \textmd{ if } e \in C
      \end{array}
    \right.
  \end{array}
\end{equation*}
Thus $\valueof{e}_{v, v_{\od_n}}$ is the value of $e$ on the
valuation $v$, parameter valuation $v_{\od_n}$, or constant $e$.


An \emph{assignment} $\pi$ is of the form
\begin{equation*}
  ( x_{k_1} x_{k_2} \dots x_{k_n} ) \mapsto 
  ( e_{l_1} e_{l_2} \dots e_{l_n} )
\end{equation*}
where $x_{k_i} \in X$, $e_{l_i} \in X \cup P_n \cup C$, and $x_{k_i}\neq x_{k_j}$
whenever $i \neq j$. Let $\Pi$ denote the set of assignments. For
valuation $v$ and parameter valuation $v_{\od_n}$,
define 

\begin{equation*}
  \valueof{\pi}_{v, v_{\od_n}} \defn
  \{ v' | v' (x_{k_i}) = \valueof{e_{l_i}}_{v, v_{\od_n}} \textmd{ for
  every } 1 \leq i \leq n \}.
\end{equation*}
That is, $\valueof{\pi}_{v, v_{\od_n}}$ contains the valuations
obtained by executing the assignment under the valuation $v$ and
parameter valuation $v_{\od_n}$.


An \emph{atomic guard} is of the form $e = f$ or its negation 
$\neg (e = f)$ (written $e \neq f$) where $e, f \in X \cup P_n \cup C$. 
A \emph{guard} is a conjunction of
atomic guards. We 
write $\Gamma$ for the set of guards. For
any valuation $v$ and parameter valuation $v_{\od_n}$, define
\begin{equation*}
  \begin{array}{ll}
    v, v_{\od_n} \models e = f & \textmd{ if } 
    \valueof{e}_{v, v_{\od_n}} = \valueof{f}_{v, v_{\od_n}} \\
    v, v_{\od_n} \models e \neq f & \textmd{ if } 
    \valueof{e}_{v, v_{\od_n}} \neq \valueof{f}_{v, v_{\od_n}} \\
    v, v_{\od_n} \models g_1 \wedge g_2 \wedge \cdots \wedge g_k &
    \textmd{ if }
    v, v_{\od_n} \models g_i \textmd{ for every } 1 \leq i \leq k
  \end{array}
\end{equation*}

\hide{
Let $g$ be a guard and $r$ an atomic guard. We say $g$ \emph{implies}
$r$ (written $\models g \implies r$) if there is no valuation $v$ and
parameter valuation $v_{\od_n}$ such that $v, v_{\od_n} \models g
\wedge \neg r$.
}

\begin{definition}
  A \emph{register automaton} is a tuple $(\Sigma, A, X, L, l_{0},
  \Delta)$ where 
  \begin{itemize}
  \item $A$ is a finite set of actions;
  \item $L$ is a finite set of \emph{locations};
  \item $l_{0} \in L$ is the \emph{initial location};
  \item $X$ is a finite set of registers.
  \item $\Delta \subseteq L \times A \times \Gamma \times \Pi \times L$ is
    a finite set of \emph{transitions}. 
\hide{
In a transition $(l, \alpha,
    g, \pi, l')$, $l$ and $l'$ are the \emph{source} and \emph{target}
    locations respectively.
}
\end{itemize}
\end{definition}

A \emph{configuration} $\langle l, v \rangle$ of a register automaton $(
\Sigma, A, X, L, l_0, \Delta)$ consists of a location $l \in L$ and a
valuation $v \in V_{(X,\Sigma)}$. For configurations 
$\langle l, v \rangle$ and $\langle l', v' \rangle$, we say $\langle
l, v \rangle$ \emph{transits} to $\langle l', v' \rangle$ on $\alpha
(\od_n)$ (written
$\langle l, v \rangle \xrightarrow{\alpha(\od_n)} \langle l', v'\rangle$) if 
there is a transition $(l, \alpha, g, \pi, l') \in \Delta$ such that 
$v, v_{\od_n} \models g$ and $v' \in \valueof{\pi}_{v,v_{\od_n}}$.

A \emph{run} of a register automaton $(\Sigma, A, X, L, l_0, \Delta)$ on
a string $\alpha_0(\od^0_{n_0})$ $\alpha_1(\od^1_{n_1}) \cdots$
$\alpha_{k-1}(\od^{k-1}_{n_{k-1}})$ is a sequence of configurations 
$\langle l_0, v_0 \rangle$ $\langle l_1, v_1\rangle \cdots$
$\langle l_k, v_k \rangle$ such that $\langle l_i, v_i \rangle
\xrightarrow{\alpha(\od^{i}_{n_{i}})} \langle l_{i+1}, v_{i+1} \rangle$ for
every $0 \leq i < k$.

\begin{figure}[htb]
\begin{center}
\begin{tikzpicture}[->,>=stealth',shorten >=1pt,auto,
  node distance=.48\textwidth]

  \node[initial,state] (0) {$l_0$};
  \node[state] (1) [right of=0] {$l_1$};

  \path[every node/.style={font=\sffamily\small}]
    (0) 
        edge [bend left] node[above] 
        {$\dfrac{\alpha (p_1, p_2)|p_1\neq p_2}{x_1\mapsto p_1,x_2\mapsto p_2}$}
        (1)
        edge [loop above] node 
        {$\dfrac{\alpha (p_1, p_2)|p_1=p_2}{-}$}
        (0)
    (1)
        edge [bend left] node[above] 
        {$\dfrac{\beta (p_1)|x_1\neq p_1 \wedge x_2\neq p_1 \wedge p_1\neq 2}{-}$}
        (0)
        edge [loop right] node[right] 
        {$\dfrac{\beta (p_1)|x_1=p_1}{\ \ x_1\mapsto x_1, x_2\mapsto x_2}$}
        (1)
        edge [loop above] node[above] 
        {$\dfrac{\beta (p_1)|x_2=p_1}{x_1\mapsto x_1, x_2\mapsto x_2}$}
        (1)
        edge [loop below] node[right] 
        {$\dfrac{\beta (p_1)|p_1=2}{x_1\mapsto p_1, x_2 \mapsto x_2}$}
        (1)
        ;
\end{tikzpicture}

\caption{A Register Automaton}
\label{figure:ra}

\end{center}
\end{figure}
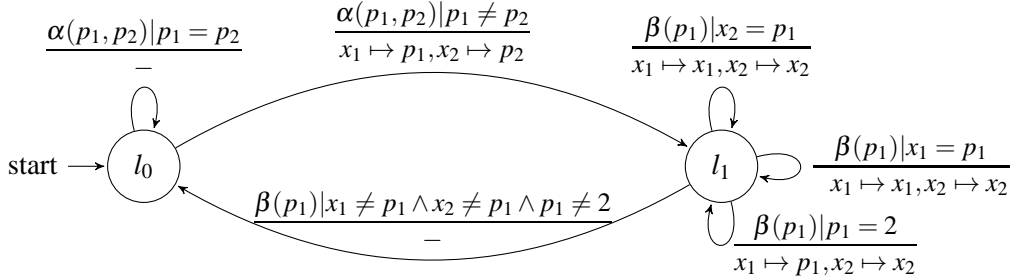

\begin{example}
Let $\bbfN$ denote the set of natural numbers, $\Sigma = \bbfN$, $A =
\{ \alpha, \beta \}$, $L = \{ l_0, l_1 \}$, $C = \{ 2 \}$, 
and $X = \{ x_1, x_2 \}$ where $\alpha$ and $\beta$ have arities 2 and
1 respectively. Consider the register automaton in
Figure~\ref{figure:ra}. In the figure, $\dfrac{\alpha |
  g}{\pi}$ denotes a transition with action $\alpha$, guard $g$, and
assignment $\pi$. Here is a run of the automaton:

\begin{equation*}
\langle l_0, 77 \rangle \xrightarrow{\alpha(1,3)}
\langle l_1, 13 \rangle \xrightarrow{\beta(1)}
\langle l_1, 13 \rangle \xrightarrow{\beta(2)}
\langle l_1, 23 \rangle \xrightarrow{\beta(1)}
\langle l_0, 69 \rangle
\end{equation*}
\label{example:ra-run}
\end{example}

Let $(\Sigma, A, X, L, l_0, \Delta)$ be a register automaton.
A configuration $\langle l, v \rangle$ is \emph{reachable} if there is
a run $\langle l_0, v_0 \rangle \langle l_1, v_1 \rangle \cdots
\langle l_k, v_k \rangle$ of $(\Sigma, A, X, L, l_0, \Delta)$ with
$\langle l_k, v_k \rangle = \langle l, v \rangle$.
The \emph{reachability} problem for register
automata is to decide whether a given configuration is reachable in a
given register automaton.


\begin{definition}[\cite{KS:08:DP}]
  An \emph{equality logic formula} is defined as follows.
  \begin{equation*}
    \begin{array}{rcl}
      \phi & : & \phi \wedge \phi\ |\ \neg \phi\ |\ \phi \implies \phi
      \ |\ \mathit{var} = \mathit{var}\\
      \mathit{var} & : & x\ |\ x' \ |\ p \ |\ c
    \end{array}
  \end{equation*}
  where $x \in X$, $x' \in X'$, $p \in P$, and $c \in C$.
\end{definition}

Note that a guard is also an equality logic formula. An equality logic
formula $\phi$ is \emph{valid} if $\phi$ always evaluates to true by
assigning each member of $X \cup X' \cup P$ with an arbitrary element in
$\Sigma$. We write $\vdash \phi$ when $\phi$ is valid. The formula
$\phi$ is \emph{consistent} if it is not the case that $\vdash \neg \phi$.
Given an equality logic formula $\phi$, the \emph{validity}
problem for equality logic is to decide whether $\vdash \phi$.


\begin{theorem}[\cite{KS:08:DP}]
  The validity problem for equality logic is coNP-complete.
\end{theorem}

%
%

\section{Representative Configurations}
\label{section:representative-configurations}

Consider a register automaton $(\Sigma, A, X, L, l_0, \Delta)$. Since
$\Sigma$ is infinite, there are an infinite number of valuations in
$V_{(X, \Sigma)}$. A register automaton subsequently has infinitely
many configurations. In this section, we show that 
configurations can be partitioned into finitely many classes. Any two
configurations in the same class are indistinguishable by register
automata.

\hide{
In this section, our goal is to give an algorithm of reachability of
configurations. Since the alphabet is infinite, there are infinite
many configurations. Because a guard  is a conjunction of equalities
or inequalities, we just care about the relations between any two
registers but the values of registers. Therefore, we define
representative configurations and show all of the behaviors of
configurations are the same as representative configurations. 
}

\hide{
In order to motivate our exposition, recall that guards are
conjunctive equality logic formulae. Since no constant is allowed in
guards, the concrete values of registers and parameters are
insignificant. Whether two registers are equal in a valuation
however is distinguishable by guards. Hence we would like to identify
two valuations with the same equality relations among registers. Since
equality relations are preserved by automorphisms, we give the
following definition.
}


\begin{definition}
  Let $u, v \in V_{(X, \Sigma)}$. $u$ is \emph{equivalent} to $v$ with respect to $C$
  (written $u \sim_C v$) if there is an automorphism $\sigma$ on $\Sigma$ 
  such that $\sigma$ is invariant on $C$ and $(\sigma \circ u)(x)= v (x)$ for every $x \in X$. 
\end{definition}

For example, let $\Sigma = \bbfN$, $X = \{ x_1, x_2, x_3 \}$, $C = \{ 1 \}$, $v_1 = 123$,
$v_2 = 134$, and $v_3 = 523$. We have $v_1\sim_C v_2$ but $v_1 \not\sim_C v_3$. 

It is easy to see that $\sim_C$ is an equivalence relation on $V_{(X,
\Sigma)}$. For any valuation $v \in V_{(X, \Sigma)}$, we write $[v]$
for the equivalence class of $v$. That is,
\begin{equation*}
  [v] \defn \{ u \in V_{(X,\Sigma)} | u \sim_C v \}.
\end{equation*}
The equivalence class $[v]$ is called a \emph{representative
  valuation}. Note that there are only finitely many representative
valuations for $X$ is finite.

\begin{definition}
  A \emph{representative configuration} $\langle l, [v]
  \rangle$ is a pair where $l \in L$ and $[v]$ is a representative
  valuation. 
\end{definition}

Since $X$ and $L$ are finite sets, the number of representative
configurations is finite. Our next task is to show that 
every configurations in a representative configuration behave
similarly. 
Let $\langle l, [v] \rangle$ and $\langle l', [v'] \rangle$ be two
representative configurations. Define
$\langle l, [v] \rangle \rightsquigarrow \langle l', [v'] \rangle$ if 
\begin{itemize}
\item for each $u \in [v]$, there is a valuation $u' \in [v']$ and a
  data symbol $\alpha(\od_n)$ such that 
  $\langle l, u \rangle \xrightarrow{\alpha(\od_n)}
  \langle l', u' \rangle$; and
\item for each $u' \in [v']$, there is a valuation $u \in [v]$ and a
  data symbol $\alpha(\od_n)$ such that 
  $\langle l, u \rangle \xrightarrow{\alpha(\od_n)}
  \langle l', u'\rangle$. 
\end{itemize}

Let $\langle \Sigma, A, X, L, l_0, \Delta \rangle$ be a register
automaton and $\langle l_k, [v_k] \rangle$ a representative
configuration. We say $\langle l_k, [v_k] \rangle$ is \emph{reachable} if
there is a sequence of representative configurations $\langle l_0, [v_0]
\rangle$ $\langle l_1, [v_1] \rangle \cdots \langle l_k, [v_k]
\rangle$ such that $\langle l_i, [v_i] \rangle \rightsquigarrow
\langle l_{i+1}, [v_{i+1}] \rangle$ for every $0 \leq i < k$.
The following three propositions are useful to our key
lemma.

\begin{proposition}
Let $v \in V_{(X, \Sigma)}$ be a valuation, $v_{\od_n} \in V_{(P,
  \Sigma)}$ a parameter valuation, and $g \in \Gamma$ a guard.
$v, v_{\od_n} \models g$ if and only if 
$\sigma \circ v, \sigma \circ v_{\od_n} \models
g$ for every automorphism $\sigma$ on $\Sigma$ which is invariant on $C$. 
\end{proposition}

\begin{proposition}
Let $v, w \in V_{(X, \Sigma)}$ be valuations, $v_{\od_n} \in V_{(P,
  \Sigma)}$ a parameter valuation, and $\pi \in \Pi$ an assignment.
$w \in \valueof{\pi}_{v, v_{\od_n}}$ if and only if 
$\sigma \circ w \in 
\valueof{\pi}_{\sigma \circ v,\sigma \circ v_{\od_n}}$ for every
automorphism $\sigma$ on $\Sigma$ which is invariant on $C$.  
\end{proposition}

\begin{proposition}
Let $(\Sigma, A, X, L, l_0, \Delta)$ be a register automaton, $l, l'
\in L$ locations, $v, v' \in V_{(X, \Sigma)}$ valuations, and $\alpha
(\od_n)$ a data symbol with $\od_n = d_1d_2\cdots d_n$.
If $\langle l, v \rangle \xrightarrow{\alpha(\od_n)}
\langle l', v' \rangle$, then $\langle l, \sigma \circ v \rangle
\xrightarrow{\alpha(\sigma(\od_n))} 
\langle l', \sigma \circ v' \rangle$ for every automorphism $\sigma$ on
$\Sigma$ which is invariant on $C$, where $\sigma (\od_n) \defn \sigma (d_1) \sigma (d_2) \cdots
\sigma (d_n)$. 
\label{proposition:representation}
\end{proposition}

\hide{
Proposition 1 and proposition 2 are from the definition of guards and
the definition of assignments respectively  and drive proposition
3.}
By Proposition~\ref{proposition:representation}, we get the following
key lemma.  

\begin{lemma}
Let $(\Sigma, A, X, L, l_0, \Delta)$ be a register automaton, $l, l'
\in L$, and $v, v' \in V_{(X, \Sigma)}$.
$\langle l, v \rangle \xrightarrow{\alpha(\od_n)}
\langle l', v' \rangle$ for some $\alpha(\od_n)$ 
if and only if
$\langle l, [v] \rangle \rightsquigarrow \langle l', [v']\rangle$.
\label{lemma:key}
\end{lemma}

\hide{
By the lemma, there is enough information in the representative
configurations for understanding how a register automaton moves. Since
the set $X$ is finite, there is only a finite number of different
representative valuations. Location set $L$ is also finite, therefore,
there is a finite number of representative configurations. From these
things, the reachability problem is become to determine whether
$\langle l, [v] \rangle \rightsquigarrow \langle l', [v'] \rangle$. 
}

\hide{
We define representative matrices to take the place of representative
valuations. Similarly, if a representative matrix couples with a
location, then it takes the place of a representative configuration. 
}

Lemma~\ref{lemma:key} shows that representative configurations are
exact representations for configurations with respect to
transitions. The configuration $\langle l, v \rangle$ transits to
another configuration $\langle l', v' \rangle$ in one step precisely
when their representative configurations have a transition. 
There are however infinitely many valuations. 
In order to enumerate $[v]$ effectively, we use a matrix-based representation.

Let $[v]$ be a representative valuation with $v \in V_{(X,
  \Sigma)}$. Assume $\{ \bar{0}, \bar{1} \} \cap \Sigma = \emptyset$. A \emph{representative matrix} $R_{[v]} \in (\{ \bar{0}, \bar{1} \} \cup C)_{|X| \times |X|}$ of $[v]$ is defined as follows.
\begin{equation*}
  \begin{array}{rcl}
  (R_{[v]})_{ij} & \defn & 
  \left\{
    \begin{array}{ll}
      v(x_i) & \textmd{ if } v(x_i) = v_(x_j) \in C\\      
      \bar{1} & \textmd{ if } v(x_{i}) = v(x_{j}) \not\in C\\      
      \bar{0} & \textmd{ otherwise}
    \end{array}
  \right.
  \end{array}
\end{equation*}

\hide{
By the definition, we make the following observation:
\begin{proposition}
  Let $R_{[v]}$ be the representative matrix of $[v]$ with $v \in
  V_{(X, \Sigma)}$. 
  \begin{enumerate}
  \item $(R_{[v]})_{ii} \in \{ \bar{1} \} \cup C$ for every $1 \leq i \leq |X|$;
  \item Symmetric: $R_{[v]}$ is symmetric; and
  \item Transitive: $R_{ij} = R_{jk} \neq \bar{0}$ implies $R_{ik} = R_{ij} = R_{jk} \neq \bar{0}$.
  \end{enumerate}
\end{proposition}
}

Let $v \in V_{(X, \Sigma)}$ be a valuation. The entry $(R_{[v]})_{ij}$ denotes the equality relation among registers $x_i$, $x_j$, and constant $c$ for every $c \in C$. If $v (x_i) = v (x_j)$, $(R_{[v]})_{ij} \in \{ \bar{1} \} \cup C$; otherwise, $(R_{[v]})_{ij} = \bar{0}$; moreover, if $v(x_i) = c \in C$, $(R_{[v]})_{ii} = c$. The following proposition shows that $R_{[v]}$ is well-defined.

\begin{proposition}
  For any $u, v \in V_{(X, \Sigma)}$, $[u] = [v]$ if and only if
  $R_{[u]} = R_{[v]}$.
  \label{proposition:r-well-defined}
\end{proposition}
\whenlongversion{
\begin{proof}
Assume $[u]=[v]$. $[u]=[v]$ if and only if $u\sim_C v$. There is an automorphism $\sigma$ which is invariant on $C$ such that $u=\sigma\circ v$. Therefore, (1) $u(x) \in C$ if and only if $\sigma(v(x)) \in C$. (2) $u(x)=u(y)$ if and only if $\sigma(v(x))=\sigma(v(y))$ if and only if $v(x)=v(y)$ for all $x,y\in X$. (3) Similarly, $u(x)\neq u(y)$ if and only if $v(x)\neq v(y)$ for all $x,y\in X$. Hence, $(R_{[u]})_{ij}=(R_{[v]})_{ij}$ for each $i,j$.
\qed
\end{proof}
}

By Proposition~\ref{proposition:r-well-defined}, we will also call $R_{[v]}$
a representative valuation and write
$R_{[v]}$ for $[v]$. Subsequently, 
$\langle l, R_{[v]} \rangle
\rightsquigarrow
\langle l', R_{[v']} \rangle$ if and only if
$\langle l, [v] \rangle
\rightsquigarrow
\langle l', [v'] \rangle$. 

\begin{example}
By example~\ref{example:ra-run}, we have
$ v_0 = 77$, $v_1 = v_2 = 13$, $v_3 = 23$, $v_4 = 69$ and  
$
R_{[v_0]}= 
\begin{pmatrix}
   \bar{1} & \bar{1}\\
   \bar{1} & \bar{1}
\end{pmatrix},
R_{[v_1]}= R_{[v_2]} = R_{[v_4]} =
\begin{pmatrix}
   \bar{1} & \bar{0}\\
   \bar{0} & \bar{1}
\end{pmatrix},
R_{[v_3]} = 
\begin{pmatrix}
    2 & \bar{0}\\
    \bar{0} & \bar{1}
\end{pmatrix}
$.
Hence, 
$\langle l,R_{[v_0]}\rangle\rightsquigarrow
\langle l',R_{[v_1]}\rangle\rightsquigarrow
\langle l',R_{[v_2]}\rangle \rightsquigarrow
\langle l',R_{[v_3]}\rangle \rightsquigarrow
\langle l',R_{[v_4]}\rangle$. 
\end{example}

Every representative valuation corresponds to a matrix. However, not
every matrix has a corresponding representative valuation. 
For instance, the zero matrix $( \bar{0} ) \in \{ \bar{0}, \bar{1} \}_{1 \times 1}$ does
not correspond to any representative valuation. If $(\bar{0}) = R_{[v]}$ for some valuation $v$, one would have the absurdity $v(x_1) \neq v(x_1)$. Such matrices are certainly not of our interests and should be excluded.

For any $R \in (\{ \bar{0}, \bar{1} \} \cup C)_{|X| \times |X|}$, define the
equality logic formula $E (R)$ as follows.
\begin{equation*}
\begin{split}
E (R) \defn 
& \bigwedge_{R_{ij} \in C} (x_i = x_j \wedge x_i = R_{ij}) \wedge  \bigwedge_{R_{ij} = \bar{1} } (x_i = x_j \wedge \bigwedge_{c \in C} x_i \neq c) \wedge \bigwedge_{R_{ij} = \bar{0} } x_i \neq x_j
\end{split}
\end{equation*}
\textbf{Idea:}
If we do not add the equalities of form $x_i = c \in C$ for some $i$ or the inequalities $x_i \neq c \in C$ for some $i$ to the conjunction $E(R)$, we can not distinguish the following four kinds of matrices:\\
(1)
$ \begin{pmatrix}
   c & \bar{1} \\
   \bar{1} & c
  \end{pmatrix}$
(2)
$ \begin{pmatrix}
   c & \bar{1} \\
   c & c
  \end{pmatrix}$
(3)
$ \begin{pmatrix}
   c & c \\
   \bar{1} & c
  \end{pmatrix}$
(4)
$ \begin{pmatrix}
   c & c \\
   c & c
  \end{pmatrix}$\\
The fourth kind of matrix is the matrix we hope for.

We say the matrix $R$ is \emph{consistent} if $E (R)$ is consistent. 
It can be shown that a consistent matrix is also a representative
matrix. Indeed, Algorithm~\ref{algorithm:canonicalval} computes a
valuation $v$ such that $R_{[v]} = R$ for any consistent matrix $R$.

\begin{algorithm}
  \tcp{$c_1, c_2, \ldots, c_{|X|}$ are distinct elements in $\Sigma \setminus C$}
  \KwIn{$R$ : a consistent matrix}
  \KwOut{$w \in V_{(X, \Sigma)}$ : $R = R_{[w]}$}
  \ForEach{$1 \leq i \leq |X|$}
  {
    \eIf{$R_{ii} \in C$}
    {
     $w (x_i) \leftarrow R_{ii}$\;
    }{
    $w (x_i) \leftarrow c_i$\;
    }
  }
  \ForEach{$i = 1$ \KwTo $|X| - 1$}
  {
    \ForEach{$j = i + 1$ \KwTo $|X|$}
    {
      \lIf{$R_{ij} \in \{ \bar{1} \} \cup C$}
      {
        $w (x_j) \leftarrow w (x_i)$\;
      }
    }
  }
  \Return $w$\;
  \caption{$\mathsf{CanonicalVal} (R)$}
  \label{algorithm:canonicalval}
\end{algorithm}

Algorithm~\ref{algorithm:canonicalval} starts from a valuation where
the register $x_i$ is assigned to $R_{ii}$ for every $R_{ii} \in C$, the rest of registers  are assigned to distinct elements in $\Sigma \setminus C$. It goes
through entries of the given consistent matrix $R$ by rows. At row
$i$, the algorithm assigns $w (x_i)$ to the register $x_j$ if $R_{ij} \in \{ \bar{1} \} \cup C$. Hence the first $i$ rows of $R$ are equal to the first $i$ rows of
$R_{[w]}$ after iteration $i$. When
Algorithm~\ref{algorithm:canonicalval} returns, we obtain a valuation
whose representative matrix is $R$.

\begin{lemma}
  Let $R \in (\{ \bar{0}, \bar{1} \} \cup C)_{|X| \times |X|}$ be a consistent matrix and
  $w = \mathsf{CanonicalVal} (R)$. $R = R_{[w]}$.
  \label{lemma:canonicalval}
\end{lemma}

For a consistent matrix $R$, the valuation computed by
$\mathsf{CanoncalVal} (R)$ is called the \emph{canonical valuation} of
$R$. The following lemma follows from Lemma~\ref{lemma:canonicalval}.

\begin{lemma}
  Let $R \in (\{ \bar{0}, \bar{1} \} \cup C)_{|X| \times |X|}$.
  $R$ is consistent if and only if $R = R_{[v]}$ for some $v \in
  V_{(X, \Sigma)}$.
  \label{lemma:universe}
\end{lemma}

By Lemma~\ref{lemma:universe}, it is now straightforward to enumerate
all representative matrices. Algorithm~\ref{algorithm:universe}
computes the set of all representative matrices.

\begin{algorithm}
  \KwOut{$\mathcal{R}$ : $\mathcal{R} = \{ R_{[v]} : v \in V_{(X, \Sigma)} \}$}
  $\mathcal{R} \leftarrow \emptyset$\;
  \ForEach{matrix $R \in (\{ \bar{0}, \bar{1} \} \cup C)_{|X| \times |X|}$}
  {
    \lIf{$R$ is consistent}
    {
      $\mathcal{R} \leftarrow \mathcal{R} \cup \{ R \}$\;
    }
  }
  \Return $\mathcal{R}$\;
  \caption{$\mathsf{UniverseR} (X)$}
  \label{algorithm:universe}
\end{algorithm}

%
%
\section{Reachability}
\label{section:reachability}

Let $(\Sigma, A, X, L, l_0, \Delta)$ be a register automaton and
$\langle l, v \rangle$ a configuration with $l \in L$ and $v \in
V_{(X, \Sigma)}$.
 In order to solve the reachability problem for
register automata, we show how to compute all $\langle l', R_{[v']}
\rangle$ such that $\langle l, R_{[v]} \rangle \rightsquigarrow
\langle l', R_{[v']} \rangle$.

By Lemma~\ref{lemma:key}, $\langle l, R_{[v]} \rangle \rightsquigarrow
\langle l', R_{[v']} \rangle$ if $\langle l, v \rangle
\xrightarrow{\alpha (\od_n)} \langle l', v' \rangle$ for some $\alpha
(\od_n)$. 
A first attempt to find $\langle l', R_{[v']} \rangle$ with $\langle
l, R_{[v]} \rangle \rightsquigarrow \langle l', R_{[v']} \rangle$ is
to compute all $\langle l', v' \rangle$ with $\langle l, v \rangle
\xrightarrow{\alpha (\od_n)} \langle l', v' \rangle$ for some $\alpha
(\od_n)$. The intuition however would not work. Since
$\Sigma$ is infinite, there can be infinitely many data symbols $\alpha
(\od_n)$ and valuations $v'$ with $\langle l, v \rangle
\xrightarrow{\alpha (\od_n)} \langle l', v' \rangle$. It is impossible
to enumerate them.

Instead, we compute $\langle l', R' \rangle$ with $\langle l, R_{[v]}
\rangle \rightsquigarrow \langle l', R' \rangle$ directly. Based on
equality relations among registers in the given configuration $\langle
l, v \rangle$, we infer equality relations among registers in a
configuration $\langle l', v' \rangle$ with $\langle l, v \rangle
\xrightarrow{\alpha (\od_n)} \langle l', v' \rangle$. Since there are
finitely many representative matrices, we enumerate those
representative matrices conforming to the inferred equality relations
among registers. The conforming representative matrices give
desired representative configurations.


We start with extracting equality relations among registers in the
given configuration $\langle l, v \rangle$. For any valuation $v \in
V_{(X, \Sigma)}$, define 
\begin{equation*}
  E (v) \defn \bigwedge_{v(x) = c \in C} x = c \wedge
         \bigwedge_{v(x) = v(y)} x = y \wedge
         \bigwedge_{v(x) \neq v(y)} x \neq y, \textmd{ and }
\end{equation*}
\begin{equation*}
  E'(v) \defn \bigwedge_{v(x) = c \in C} x' = c \wedge
         \bigwedge_{v(x) = v(y)} x' = y' \wedge
         \bigwedge_{v(x) \neq v(y)} x' \neq y'.
\end{equation*}

\hide{
Let $(l, \alpha, g, \pi, l')$ be a transition with 
the guard $g = g_1 \wedge g_2 \wedge \cdots g_k$. Define 
\begin{equation*}
  g\downarrow_X \defn \bigwedge
  \{ g_i | g_i \textmd{ is } x = y \textmd{ or } x \neq y \textmd{
    with } x, y \in X \textmd{ and } 1 \leq i \leq k \}.
\end{equation*}
That is, $g\downarrow_X$ consists of the equalities and disequalities
between registers in $g$.
}


Let $(l, \alpha, g, \pi, l')$ be a transition and $\langle l, v
\rangle \xrightarrow{\alpha (\od_n)} \langle l', v' \rangle$. Equality
relations among registers in $\langle l', v' \rangle$ are
determined by the assignment $\pi$. Let $\pi = (x_{k_1} x_{k_2} \dots
x_{k_n}) \mapsto (e_{l_1} e_{l_2} \dots e_{l_n})$. Define
\begin{equation*}
  E (\pi) \defn \bigwedge\limits^n_{i=1} x'_{k_i} = e_{l_i}.
\end{equation*}

Observe that $E (v)$ and $E (\pi)$ are equality logic formulae for any
valuation $v$ and assignment $\pi$. By Lemma~\ref{lemma:universe},
$\langle l, R \rangle$ is a representative configuration when $R$ is a
consistent matrix. For any representative
configuration $\langle l, R \rangle$, we characterize a representative
configuration $\langle l', R' \rangle$ with $\langle l, R \rangle
\rightsquigarrow \langle l', R' \rangle$ as follows.


\begin{definition}
  Let $RA = (\Sigma, A, X, L, l_0, \Delta)$ be a register
  automaton, $(l,$ $\alpha,$ $g,$ $\pi,$ $l') \in \Delta$ a
  transition, and $R$ a consistent matrix. Define the set $\Post_{RA}
  (\langle l, R \rangle)$ of  
  representative matrices as follows. $\langle l', R' \rangle
  \in \Post_{RA} (\langle l, R \rangle)$ if $g \wedge E (w) \wedge E
  (\pi) \wedge E' (w')$ is consistent,
  where $w$ and $w'$ are the canonical valuations of $R$ and $R'$ respectively.
  \label{definition:post}
\end{definition}


\begin{example}
Let $\Sigma = \bbfN$, $X = \{ x_1, x_2, x_3 \}$, and 
$R = 
\begin{pmatrix}
   \bar{1} & \bar{0} & \bar{0} \\
   \bar{0} & \bar{1} & \bar{1} \\
   \bar{0} & \bar{1} & \bar{1}
\end{pmatrix}$. By Algorithm~\ref{algorithm:canonicalval}, $w = 122$
is the canonical valuation of $R$.
Consider a transition $(l, \alpha, g, \pi, l')$ where $g$ is 
$(x_1 \neq x_2) \wedge (p_1 \neq p_2)$ and $\pi$ is
$(x_1 x_2 x_3) \mapsto (x_2 p_1 p_2)$. Then
$E (v)$ is $(x_1 \neq x_2) \wedge (x_1 \neq x_3) \wedge (x_2 = x_3)$
and $E (\pi)$ is $(x'_1 = x_2) \wedge (x'_2 = p_1) \wedge (x'_3 =
p_2)$. Let $F$ denote the equality logic formula $g \wedge E (v)
\wedge E (\pi)$. $F$ is consistent. Observe that $\vdash F \implies
x'_2 \neq x'_3$. Consider the following three cases:
\begin{enumerate}
\item $R'_0$ is
$\begin{pmatrix}
   \bar{1} & \bar{1} & \bar{0} \\
   \bar{1} & \bar{1} & \bar{0} \\
   \bar{0} & \bar{0} & \bar{1}
\end{pmatrix}$. Since $\vdash F \implies
x'_1 = x'_2 \wedge x'_1 \neq x'_3$, $\langle l', R'_0 \rangle \in
\Post_{RA} (\langle l, R \rangle)$;
\item $R'_1$ is
$\begin{pmatrix}
   \bar{1} & \bar{0} & \bar{1} \\
   \bar{0} & \bar{1} & \bar{0} \\
   \bar{1} & \bar{0} & \bar{1}
\end{pmatrix}$. Since $\vdash F \implies
x'_1 = x'_3 \wedge x'_1 \neq x'_2$, $\langle l', R'_1 \rangle \in
\Post_{RA} (\langle l, R \rangle)$;
\item $R'_2$ is
$\begin{pmatrix}
   \bar{1} & \bar{0} & \bar{0} \\
   \bar{0} & \bar{1} & \bar{0} \\
   \bar{0} & \bar{0} & \bar{1}
\end{pmatrix}$. Since $\vdash F \implies
x'_1 \neq x'_2 \wedge x'_1 \neq x'_3$, $\langle l', R'_2 \rangle \in
\Post_{RA} (\langle l, R \rangle)$.
\end{enumerate}
\end{example}


\begin{lemma}
$E(v)$ is consistent for every $v \in V_{(X,\Sigma)}$. Moreover, $E(v) = E(w)$ if $[v] = [w]$.
\label{lemma:universe'}
\end{lemma}
\whenlongversion{
\begin{proof}
Lemma~\ref{lemma:universe'} is directly from Lemma~\ref{lemma:universe}.
\qed
\end{proof}
}


\begin{lemma}
Let $RA = (\Sigma, A, X, L, l_0, \Delta)$ be a register automaton, $R$ and $R'$ be consistent. $\langle l, R \rangle \rightsquigarrow \langle l', R' \rangle$ iff there is $(l, \alpha, g, \pi, l') \in \Delta$ and $g \wedge E(w) \wedge E(\pi) \wedge E'(w')$ is consistent, where $w$ and $w'$ are the canonical valuations of $R$ and $R'$ respectively.
\label{lemma:consistent}
\end{lemma}
\whenlongversion{
\begin{proof}
$(\Rightarrow)$
Since $R$ and $R'$ are consistent, there $w$, $w' \in V_{(X,\Sigma)}$ such that $R = R_{[w]}$ and $R' = R_{[w']}$. Therefore, $\langle l, R \rangle \rightsquigarrow \langle l', R' \rangle$ iff $\langle l, [w] \rangle \rightsquigarrow \langle l', [w'] \rangle$ iff there is $(l, \alpha, g, \pi, l') \in \Delta$, $v \in [w]$, $v' \in [w']$, and $\alpha(\od_n)$ such that $\langle l, v \rangle  \xrightarrow{\alpha(\od_n)} \langle l', v' \rangle$. By Lemma~\ref{lemma:universe'}, $g \wedge E(w) \wedge E(\pi) \wedge E(w') = g \wedge E(v) \wedge E(\pi) \wedge E(v')$. Since $v, v_{\od_n} \models g$ and $v' \in \valueof{\pi}_{v,v_{\od_n}}$, $g \wedge E(v) \wedge E(\pi) \wedge E(v')$ is consistent.\\
$(\Leftarrow)$
Since $g \wedge E(w) \wedge E(\pi) \wedge E'(w')$ is consistent, there are $v \in [w]$, $v' \in [w']$, and $v_{\od_n} \in \Sigma^*$ such that $v, v_{\od_n} \models g$, $v' \in \valueof{\pi}_{v,v_{\od_n}}$, and $\langle l, v \rangle  \xrightarrow{\alpha(\od_n)} \langle l', v' \rangle$.
\qed
\end{proof}
}


The following lemma is directly from Lemma~\ref{lemma:consistent}. It shows that Definition~\ref{definition:post}
correctly characterizes successors of any given representative
configuration.
\begin{lemma}
$\Post_{RA} (\langle l, R \rangle)$ $=$ $\{ \langle
l', R' \rangle | \langle l, R \rangle \rightsquigarrow \langle l', R'
\rangle \}$.
\label{lemma:post}
\end{lemma}

Using Algorithm~\ref{algorithm:universe} to enumerate representative
matrices, it is straightforward to compute the set $\Post_{RA} (\langle l,
R \rangle)$ for any representative configuration $\langle l, R \rangle$ 
(Algorithm~\ref{algorithm:post}). We first obtain the canonical
valuation $w$ for $R$. The algorithm iterates through transitions of
the given register automaton. For a transition $(l, \alpha, g, \pi,
l')$, define the equality logic formula $F$ to be $g \wedge E (w)
\wedge E (\pi)$. The algorithm then checks if $F$ is consistent. 
If so, it goes through every representative matrices and adds them to
the successor set $U$ by Lemma~\ref{lemma:post}. 

\begin{algorithm}
  \KwIn{$RA$: $(\Sigma, A, X, L, l_0, \Delta)$;
    $\langle l, R \rangle$ : a representative configuration}
  $\mathcal{R} \leftarrow \mathsf{UniverseR} (X)$\;
  $U, w \leftarrow \emptyset, \mathsf{CanonicalVal} (R)$\;
  \ForEach {$(l, \alpha, g, \pi, l') \in \Delta$}
  {
    $F \leftarrow g \wedge E (w) \wedge E (\pi)$\;
    \uIf{$F$ is consistent}
    {
      \ForEach {$R' \in \mathcal{R}$}
      {
        $w' \leftarrow \mathsf{CanonicalVal} (R')$\;
        $F' \leftarrow g \wedge E (w) \wedge E (\pi) \wedge E' (w')$\;
        \lIf{$F'$ is consistent}
        {
          $U \leftarrow U \cup \{ R' \}$\;
        }
      }
    }
  }
  \Return $U$\;
  \caption{$\mathsf{Post} (RA, \langle l, R \rangle)$}
  \label{algorithm:post}
\end{algorithm}

\begin{theorem}
  Let $RA = (\Sigma, A, X, L, l_0, \Delta)$ be a register automaton
  and $\langle l, R \rangle$ a representative configuration.
  $R' \in \Post_{RA} (\langle
  l, R \rangle)$ iff $R' \in \mathsf{Post} (RA, \langle l,
  R \rangle)$.
  \label{theorem:post}
\end{theorem}

With the algorithm $\mathsf{Post} (RA, \langle l, R \rangle)$ at hand, we
are ready to present our solution to the reachability problem for
register automata. By Lemma~\ref{lemma:key}, $\langle l_0, v_0 \rangle
\langle l_1, v_1 \rangle \cdots \langle l_k, v_k \rangle$ is a run
precisely when $\langle l_0, R_{[v_0]} \rangle \rightsquigarrow
\langle l_1, R_{[v_1]} \rangle \rightsquigarrow \cdots \rightsquigarrow
\langle l_k, R_{[v_k]} \rangle$. In order to check if the
configuration $\langle l, v \rangle$ is reachable, we compute
reachable representative configurations and check if the
$\langle l, R_{[v]} \rangle$ belongs to the reachable representative
configurations (Algorithm~\ref{algorithm:reach}).

\begin{algorithm}[h]
  \KwIn{$(\Sigma, A, X, L, l_0, \Delta)$ : a register automaton;
        $\langle l, R \rangle$ : a representative configuration}
  \KwOut{$\mathit{true}$ if $\langle l, R \rangle$ is reachable;
    $\mathit{false}$ otherwise} 

  $\mathcal{R} \leftarrow \mathsf{UniverseR} (X)$\;
  $U, V \leftarrow 
  \{ \langle l_0, R_0 \rangle | R_0 \in \mathcal{R} \}, \emptyset$\;
  \While{$U \neq V$}
  {
    $U' \leftarrow \bigcup_{\langle l, R \rangle \in U}
    \mathsf{Post} (RA, \langle l, R \rangle)$\;
    $V, U \leftarrow U, U \cup U'$\;
  }
  $\mathit{result} \leftarrow$
  \lIf{$\langle l, R \rangle \in U$}
  {
    $\mathit{true}$
  }
  \lElse
  {
    $\mathit{false}$\;
  }
  \Return $\mathit{result}$\;
  \caption{$\mathsf{Reach} ((\Sigma, A, X, L, l_0, \Delta), \langle l, R\rangle)$}
  \label{algorithm:reach}
\end{algorithm}

Our first technical result is summarized in the following theorem.

\begin{theorem}
  Let $(\Sigma, A, X, L, l_0, \Delta)$ be a register automaton and
  $\langle l, v \rangle$ a configuration.
  $\langle l, v \rangle$ is
  reachable iff
  $\mathsf{Reach} ((\Sigma, A, X, L, l_0, \Delta), (l, R_{[v]}))$ returns
  $\mathit{true}$. 
\end{theorem}

%
%

\section{$\mathit{CTL} (X, L)$ Model Checking}
\label{section:model-checking}

In addition to checking whether a configuration is reachable, it is
often desirable to check patterns of configurations in runs of a
register automaton. We define a computation tree logic to specify
patterns of configurations in register automata. Representative
configurations are then used to design an algorithm that solves the
model checking problem for register automata.

Let $X$ be the set of registers and $L$ the set of locations. 
An \emph{atomic formula} is an equality over
$X$, an equality one side over $X$ another side over $C$, or a location $l \in L$. We write $AP$ for the set of atomic
formulae. Consider 
the computation tree logic $\mathit{CTL} (X, L)$ defined as
follows~\cite{R3}.
\begin{itemize}
\item If $f \in AP$, $f$ is a $\mathit{CTL} (X, L)$ formula;
\item If $f_0$ and $f_1$ are $\mathit{CTL} (X, L)$ formulae, $\neg
  f_0$ and $f_0 \wedge f_1$ are $\mathit{CTL} (X, L)$ formulae;
\item If $f_0$ and $f_1$ are $\mathit{CTL} (X, L)$ formulae, $EX f_0$,
  $E (f_0 \until f_1)$, and $EG f_0$ are $\mathit{CTL} (X, L)$ formulae.
\end{itemize}
We use the standard abbreviations: $\mathit{false} (\equiv \neg (x = x))$,
$\mathit{true} (\equiv \neg \mathit{false})$, 
$f_0 \vee f_1 (\equiv \neg (\neg f_0 \wedge \neg f_1))$, 
$f_0 \implies f_1 (\equiv \neg f_0 \vee f_1)$, $AX f_0
(\equiv \neg EX \neg f_0)$, $EF f_0 (\equiv E (\mathit{true} \until f_0))$,
$AG f_0 (\equiv \neg EF \neg f_0)$, and $AF f_0 (\equiv \neg EG \neg
f_0)$. Examples of $\mathit{CTL} (X, L)$ are $AF (l_{\mathit{end}}
\wedge x_1 = x_2)$, $AG ((l_{\mathit{start}} \wedge \neg (x_1 = x_2))
\implies EF (l_{\mathit{end}} \wedge (x_1 = x_2)))$.

Let $\langle l, v \rangle$ be a configuration of a register automaton
$RA = (\Sigma, A, X, L, l_0, \Delta)$ and $f$ a $\mathit{CTL} (X, L)$
formula. Define \emph{$\langle l, v \rangle$ satisfies $f$ in $RA$}
($\langle l, v \rangle \models_{RA} f$) by
\begin{itemize}
\item $\langle l, v \rangle \models_{RA} l$;
\item $\langle l, v \rangle \models_{RA} x = y$ if $v (x) = v (y)$;
\item $\langle l, v \rangle \models_{RA} \neg f$ if not
  $\langle l, v \rangle \models_{RA} f$;
\item $\langle l, v \rangle \models_{RA} f_0 \wedge f_1$ if
  $\langle l, v \rangle \models_{RA} f_0$ and $\langle l, v \rangle
  \models_{RA} f_1$;
\item $\langle l, v \rangle \models_{RA} EX f$ if
  $\langle l', v' \rangle \models_{RA} f$ for some $\alpha (\od_n)$
  such that $\langle l, v \rangle \xrightarrow{\alpha (\od_n)} \langle
  l', v' \rangle$;
\item $\langle l, v \rangle \models_{RA} E (f_0 \until f_1)$ if there
  are $k \geq 0$, $\alpha_i (\od^i_{n_i})$, $\langle l_i, v_i \rangle$ with 
  $\langle l_0, v_0 \rangle = \langle l, v \rangle$, and
  $\langle l_i, v_i \rangle \xrightarrow{\alpha_{i}
    (\od^{i}_{n_{i}})} \langle l_{i+1}, v_{i+1} \rangle$ for every
  $0 \leq i < k$ such that
  (1) $\langle l_k, v_k \rangle \models_{RA} f_1$; and
  (2) $\langle l_i, v_i \rangle \models_{RA} f_0$ for every $0 \leq
    i < k$.
\item $\langle l, v \rangle \models_{RA} EG f$ if there are $\alpha_i
  (\od^i_{n_i})$, $\langle l_i, v_i \rangle$ with $\langle l_0, v_0
  \rangle = \langle l, v \rangle$, and $\langle l_i, v_i
  \rangle \xrightarrow {\alpha_{i} (\od^{i}_{n_{i}})}
  \langle l_{i+1}, v_{i+1} \rangle$ for every $i \geq 0$ such that
  $\langle l_i, v_i \rangle \models_{RA} f$.
\end{itemize}

Let $RA = (\Sigma, A, X, L, l_0, \Delta)$ be a register automaton and $f$ a
$\mathit{CTL} (X, L)$ formula. We say $RA$ \emph{satisfies} $f$
(written $\models_{RA} f$) if $\langle l_0, v \rangle \models_{RA} f$
for every $v \in V_{(X, \Sigma)}$. The \emph{$\mathit{CTL} (X, L)$ model
  checking} problem for register automata is to decide whether
$\models_{RA} f$. The following lemma shows that any two
configurations in a representative configuration satisfy the same
$\mathit{CTL} (X, L)$ formulae.


\begin{lemma}
  Let $RA = (\Sigma, A, X, L, l_0, \Delta)$ be a register automaton,
  $l \in L$, $u, 
  v \in V_{(X, \Sigma)}$, and $f$ a $\mathit{CTL} (X, L)$ formula. If
  $u \sim_C v$, then
  \begin{equation*}
    \langle l, u \rangle \models_{RA} f
    \textmd{ if and only if }
    \langle l, v \rangle \models_{RA} f.
  \end{equation*}
  \label{lemma:ctl-representation}
\end{lemma}

By Lemma~\ref{lemma:ctl-representation}, it suffices to compute 
representative configurations for any $\mathit{CTL} (X, L)$
formula. For any $\mathit{CTL} (X, L)$ formula $f$, we compute the set
of representative configurations $\{ \langle l, R_{[v]} \rangle |
\langle l, v \rangle \models_{RA} f \}$. Our model checking algorithm
essentially follows the classical algorithm for finite-state
models.


\begin{algorithm}
  \KwIn{$RA$: $(\Sigma, A, X, L, l_0, \Delta)$; $ap$ : a
    $\mathit{CTL} (X, L)$ atomic formula}
  \KwOut{$\{ \langle l, R_{[v]} \rangle | \langle l, v \rangle
    \models_{RA} ap \}$}
  $\mathcal{R} \leftarrow \mathsf{UniverseU} (X)$\;
  \Switch{$ap$}
  {
    \lCase{$l$:}
    { \Return $\{ \langle l, R \rangle | R \in \mathcal{R} \}$\; }
    \lCase{$x_i = x_j$:}
    { \Return $L \times \{ R \in \mathcal{R} | R_{ij} = \bar{1} \textmd{ or } R_{ij} = c 
    \in C \}$\; }
    \lCase{$x_i = c$:}
    { \Return $L \times \{ R \in \mathcal{R} | R_{ii} = c \in C\}$\;}
  }
  \caption{$\mathsf{ComputeAP} (RA, ap)$}
  \label{algorithm:computeap}
\end{algorithm}

Algorithm~\ref{algorithm:computeap} computes the set of representative
configurations for atomic propositions. Clearly,
$\mathsf{ComputeAP} (RA, ap) = \{ \langle l, R_{[v]} \rangle | \langle
l, v \rangle \models_{RA} ap \}$.

\hide{
\subsection*{Cases: $\neg f$,$f_1\vee f_2$, and $EXf$}
\begin{itemize}
\item $\neg f$: $\{\langle l,V\rangle|\langle l,V\rangle\models_{RA}\neg f\}=\{\langle l,V\rangle|\langle l,V\rangle\models_{RA}f\}^C$.
\item $f_1\vee f_2$: $\{\langle l,V\rangle|\langle l,V\rangle\models_{RA}f_1\vee f_2\}=\{\langle l,V\rangle|\langle l,V\rangle\models_{RA}f_1\}\cup\{\langle l,V\rangle|\langle l,V\rangle\models_{RA}f_2\}$.
\item $EXf$: $\{\langle l,V\rangle|\langle l,V\rangle\models_{RA}EXf\}=\{\langle l,V\rangle|\langle l',V'\rangle\models_{RA}f,\exists \langle l',V'\rangle\in N_{\langle l,V\rangle}\}$.
\end{itemize}
}


\begin{algorithm}
  \KwIn{$RA$: $(\Sigma, A, X, L, l_0, \Delta)$;
        $S$ : $\{ \langle l, R_{[v]} \rangle | \langle l, v \rangle
        \models_{RA} f \}$}
  \KwOut{$\{ \langle l, R_{[v]} \rangle | \langle l, v \rangle
    \models_{RA} \neg f \}$}
  $\mathcal{R} \leftarrow \mathsf{UniverseR} (X)$\;
  \Return $(L \times \mathcal{R}) \setminus S$\;
  \caption{$\mathsf{ComputeNot} (RA, S)$}
  \label{algorithm:computenot}
\end{algorithm}

\begin{algorithm}
  \KwIn{$RA$ : $(\Sigma, A, X, L, l_0, \Delta)$;
        $S_0$ : $\{ \langle l, R_{[v]} \rangle | \langle l, v \rangle
        \models_{RA} f_0 \}$;
        $S_1$ : $\{ \langle l, R_{[v]} \rangle | \langle l, v \rangle
        \models_{RA} f_1 \}$}
  \KwOut{$\{ \langle l, R_{[v]} \rangle | \langle l, v \rangle
    \models_{RA} f_0 \wedge f_1 \} $}
  \Return $S_0 \cap S_1$\;
  \caption{$\mathsf{ComputeAnd} (RA, S_0, S_1)$}
  \label{algorithm:computeand}
\end{algorithm}

For Boolean operations, we assume that representative configurations
for operands have been computed. Algorithm~\ref{algorithm:computenot}
and~\ref{algorithm:computeand} give details 
for the negation and conjunction of $\mathit{CTL} (X,
L)$ formulae respectively.

\begin{algorithm}
  \KwIn{$RA$: $(\Sigma, A, X, L, l_0, \Delta)$;
    $S$ : $\{ \langle l, R_{[v]} \rangle | \langle l, v \rangle
    \models_{RA} f \}$}
  \KwOut{$\{ \langle l, R_{[v]} \rangle | \langle l, v \rangle
    \models_{RA} EX f \}$}
  $\mathcal{R}, U \leftarrow \mathsf{UniverseR} (X), \emptyset$\;
  \ForEach {$\langle l, R \rangle \in L \times \mathcal{R}$}
  {
    \lIf{$\mathsf{Post} (RA, \langle l, R \rangle) \cap S \neq \emptyset$}
    {
      $U \leftarrow U \cup \{ \langle l, R \rangle \}$\;
    }
  }
  \Return $U$\;
  \caption{$\mathsf{ComputeEX} (RA, S)$}
  \label{algorithm:computeex}
\end{algorithm}

Given the set $S$ of representative configurations for a $\mathit{CTL}
(X, L)$ formula $f$, Algorithm~\ref{algorithm:computeex} shows how to
compute representative configurations for $EX f$. For every possible
representative configuration $\langle l, R \rangle$, it checks if
$\langle l', R' \rangle \in S$ for some $\langle l', R' \rangle$ with
$\langle l, R \rangle \rightsquigarrow \langle l', R' \rangle$. If so,
$\langle l, R \rangle$ is added to the result.

\hide{
\subsection*{Case: $E[f_1\cup f_2]$}
If $\langle l^0,V^0\rangle\in\{\langle l,V\rangle|\langle l,V\rangle\models_{RA}f_2\}$, let $A_0=\{\langle l^0,V^0\rangle\}$, and $A_{i-1}=\{\langle l,V\rangle|\langle l,V\rangle\models_{RA}f_1\wedge\langle l',V'\rangle\in N_{\langle l,V\rangle},\exists\langle l',V'\in A_{i-1}\}$ for all $i\geq 1$. Then $\bigcup_iA_i\subseteq\{\langle l,V\rangle|\langle l,V\rangle\models_{RA}E[f_1\cup f_2]\}$.
}

\begin{algorithm}
  \KwIn{$RA$: $(\Sigma, A, X, L, l_0, \Delta)$;
        $S_0$ : $\{ \langle l, R_{[v]} \rangle | \langle l, v \rangle
        \models_{RA} f_0 \}$;
        $S_1$ : $\{ \langle l, R_{[v]} \rangle | \langle l, v \rangle
        \models_{RA} f_1 \}$}
  \KwOut{$\{ \langle l, R_{[v]} \rangle | \langle l, v \rangle
    \models_{RA} f_0 \until f_1 \}$}
  $U, V \leftarrow S_1, \emptyset$\;
  \While{$U \neq V$}
  {
    $W \leftarrow \mathsf{ComputeEX} (RA, U)$\;
    $V, U \leftarrow U, U \cup (W \cap S_0)$\;
  }
  \Return $U$\;
  \caption{$\mathsf{ComputeEU} (RA, S_0, S_1)$}
  \label{algorithm:computeu}
\end{algorithm}

To compute representative configurations for $f_0 \until f_1$, recall
that $f_0 \until f_1$ is the least fixed point of the function $\Psi
(Z) = f_1 \vee (f_0 \wedge EX Z)$. Algorithm~\ref{algorithm:computeu}
thus follows the standard fixed point computation for the
$\mathit{CTL} (X, L)$ formula $f_0 \until f_1$.

\hide{
\subsection*{Case: $EGf$}
If $\langle l^0,V^0\rangle\in\{\langle l,V\rangle|\langle l,V\rangle\models_{RA}f\}$, let $A_0=\{\langle l^0,V^0\rangle\}$, $A_i=\{\langle l',V'\rangle|\langle l',V'\rangle\models_{RA}f\wedge\langle l',V'\rangle\in N_{\langle l,V\rangle},\exists \langle l,V\rangle\in A_{i-1}\}\setminus A_{i-1}$ for all $i\geq 1$, and $B_i=\bigcup_{k=0}^{i-1}A_k$ for all $i\geq 1$. If $A_i\cap B_i\neq\emptyset$ for some $i$ ,then $\langle l^0,V^0\rangle\in\{\langle l,V\rangle|\langle l,V\rangle\models_{RA}EGf\}$.
}

\begin{algorithm}
  \KwIn{$RA$ : $(\Sigma, A, X, L, l_0, \Delta)$;
        $S$ : $\{ \langle l, R_{[v]} \rangle | \langle l, v \rangle
        \models_{RA} f \}$}
  \KwOut{$\{ \langle l, R_{[v]} \rangle | \langle l, v \rangle
    \models_{RA} EG f \}$}
  $U, V \leftarrow S, \mathsf{UniverseR} (X)$\;
  \While{$U \neq V$}
  {
    $W \leftarrow \mathsf{ComputeEX} (RA, U)$\;
    $V, U \leftarrow U, U \cap W$\;
  }
  \Return $U$\;
  \caption{$\mathsf{ComputeEG} (RA, S)$}
  \label{algorithm:computeeg}
\end{algorithm}

For the $\mathit{CTL} (X, L)$ formula $EG f$, recall that $EG f$ is
the greatest fixed point of the function $\Phi (Z) = f \wedge EX
Z$. Algorithm~\ref{algorithm:computeeg} performs the greatest fixed
point computation to obtain representative configurations for $EG f$.

\begin{algorithm}
  \KwIn{$RA$ : $(\Sigma, A, X, L, l_0, \Delta)$;
        $f$ : a $\mathit{CTL} (X, L)$ formula}
  \KwOut{$\{ \langle l, R_{[v]} \rangle | \langle l, v \rangle
    \models_{RA} f \}$}
  \Switch{$f$}
  {
    \uCase{$l$, $x_i = x_j$, or $x_i = c$:}
    { 
      $U \leftarrow \mathsf{ComputeAP} (RA, f)$\;
    }
    \uCase{$\neg f_0$:}
    {
      $V \leftarrow \mathsf{ComputeCTL} (RA, f_0)$\;
      $U \leftarrow \mathsf{ComputeNot} (RA, V)$\;
    }
    \uCase{$f_0 \wedge f_1$:}
    {
      $V_0, V_1 \leftarrow 
      \mathsf{ComputeCTL} (RA, f_0), 
      \mathsf{ComputeCTL} (RA, f_1)$\;
      $U \leftarrow \mathsf{ComputeAnd} (RA, V_0, V_1)$\;
    }
    \uCase{$EX f_0$:}
    {
      $V \leftarrow \mathsf{ComputeCTL} (RA, f_0)$\;
      $U \leftarrow \mathsf{ComputeEX} (RA, V)$\;
    }
    \uCase{$E (f_0 \until f_1)$:}
    {
      $V_0, V_1 \leftarrow 
      \mathsf{ComputeCTL} (RA, f_0),
      \mathsf{ComputeCTL} (RA, f_1)$\;
      $U \leftarrow \mathsf{ComputeEU} (RA, V_0, V_1)$\;
    }
    \uCase{$EG f_0$:}
    {
      $V \leftarrow \mathsf{ComputeCTL} (RA, f_0)$\;
      $U \leftarrow \mathsf{ComputeEG} (RA, V)$\;
    }
  }
  \Return $U$\;
  \caption{$\mathsf{ComputeCTL} (RA, f)$}
  \label{algorithm:computectl}
\end{algorithm}

The representative configurations for a $\mathit{CTL} (X, L)$ formula
are computed by induction on the formula
(Algorithm~\ref{algorithm:computectl}). Theorem~\ref{theorem:ctl}
summaries the algorithm.

\begin{theorem}
  Let $RA = (\Sigma, A, X, L, l_0, \Delta)$ be a register automaton,
  $f$ a $\mathit{CTL} (X, L)$ formula, $l \in L$, and $v \in V_{(X,
    \Sigma)}$. $\langle l, v \rangle \models_{RA} f$ if and only if
  $\langle l, R_{[v]} \rangle \in \mathsf{ComputeCTL} (RA, f)$.
  \label{theorem:ctl}
\end{theorem}

It is easy to check whether $\models_{RA} f$ for any register
automaton $RA$ and $\mathit{CTL} (X, L)$ formula $f$ by
Theorem~\ref{theorem:ctl} (Algorithm~\ref{algorithm:modelcheck}). 
We compute the set $\{ \langle l, R_{[v]} \rangle | \langle l, v
\rangle \models_{RA} f \}$ of representative configurations and check
if $\langle l, R \rangle$ belongs to the set for every representative
matrix $R$.

\begin{algorithm}
  \KwIn{$RA$ : $(\Sigma, A, X, L, l_0, \Delta)$;
        $f$ : a $\mathit{CTL} (X, L)$ formula}
  \KwOut{$\mathit{true}$ if $\models_{RA} f$;
         $\mathit{false}$ otherwise}
  $U \leftarrow \mathsf{ComputeCTL} (RA, f)$\;
  $\mathcal{R} \leftarrow \mathsf{UniverseR} (X)$\;
  $W \leftarrow \{ \langle l_0, R \rangle | R \in \mathcal{R} \}$\;
  $\mathit{result} \leftarrow$
  \lIf{$W \subseteq U$}
  { $\mathit{true}$ }
  \lElse
  { $\mathit{false}$\; }
  \Return $\mathit{result}$\;
  \caption{$\mathsf{ModelCheck} (RA, f)$}
  \label{algorithm:modelcheck}
\end{algorithm}

%
%

\section{An Example}
\label{section:example}

In the Byzantine generals problem, one
commanding and $n - 1$ lieutenant generals would like to share
information through one-to-one communication. However, not all
generals are loyal. Some of them (the commanding general included) may
be traitors. Traitors need not follow rules. The problem is to
devise a mechanism so that all loyal generals share the same
information at the end.

Consider the
scenario with a commanding general, two loyal lieutenant, and
one treacherous general. The emperor decides to send $m$ soldiers to
the front line, and asks the commanding general to inform
the lieutenant generals. Based on the algorithm in~\cite{W:12:ASBGP}, we
give a model where a loyal, the treacherous, and the other loyal
lieutenant generals act in turn. We want to know the initial
configurations where both loyal generals agree upon the same
information in this setting.  

Since the number of soldiers is unbounded, we choose $\bbfN$ as the
infinite alphabet. The set of constants $C$ is $\{ 0 \}$, it is for default decision.
When a lieutenant general cannot decide, he will take the default decision. 
We identify lieutenant
generals by numbers: $1$ and $2$ are loyal, $\ug$ is
treacherous. The action set $A$ has four actions: $\alpha_{1},
\alpha_{2}, \alpha_{\ug}$, and $\alpha_M$. The action $\alpha_{i}$
means that the lieutenant general $i$ receives messages from the other
lieutenant generals. Each lieutenant general computes the
majority of messages in action
$\alpha_M$. Eight registers will be used. The registers $r_{1}, r_{2}, r_{\ug}$
contain the commanding general's messages sent to each lieutenant
general respectively. The final decisions of each lieutenant generals
are stored in the registers $D_{1}, D_{2}$, and $D_{\ug}$
respectively. Finally, $s$ and $t$ are temporary registers.

\begin{figure}
  \centering
  \begin{tikzpicture}[->,>=stealth',shorten >=1pt,auto,
    node distance=.45\textwidth]
    \tikzstyle{every state}=     [minimum size=.075\textwidth]
    \node[initial, state] (2) {$l_0$};
    \node[state] (3) [right of=2] {$l_{1}$};
    \node[state] (4) [right of=3] {$L_{1}$};

    \path[every node/.style={font=\sffamily\tiny}]
    
    (2) 
    edge [left] node[above] 
    {$\dfrac{\alpha_{1}(p_1,p_2)|p_1=r_2}{
        \begin{matrix}
          (r_1,r_2,r_{\ug},s,t) \mapsto(r_1,r_2,r_{\ug},p_1,p_2)
        \end{matrix} 
      }$}
    (3)

    (3)
    edge [bend left=60] node[above] 
    {$\dfrac{\alpha_M|r_1=s}
      {\begin{matrix}
          (r_1, r_2, r_{\ug}, D_1) \mapsto (r_1, r_2, r_{\ug}, s)
        \end{matrix}}$}
    (4)

    (3)
    edge [bend left=12] node[above] 
    {$\dfrac{\alpha_M|r_1=t}
      {\begin{matrix}
          (r_1, r_2, r_{\ug}, D_1) \mapsto (r_1, r_2, r_{\ug}, t)
        \end{matrix}}$}
    (4)

    (3)
    edge [bend right=12] node[below] 
    {$\dfrac{\alpha_M|s = t}
      {\begin{matrix}
          (r_1, r_2, r_{\ug}, D_1) \mapsto (r_1, r_2, r_{\ug}, s)
        \end{matrix}}$}
    (4)
    
    (3)
    edge [bend right=60] node[below] 
    {$\dfrac{\alpha_M|\textmd{else}}
      {\begin{matrix}
          (r_1, r_2, r_{\ug}, D_1) \mapsto (r_1, r_2, r_{\ug}, 0)
        \end{matrix}}$}
    (4)
    ;
  \end{tikzpicture}
  \caption{The Lieutenant General $1$}
  \label{figure:lg1}
\end{figure}
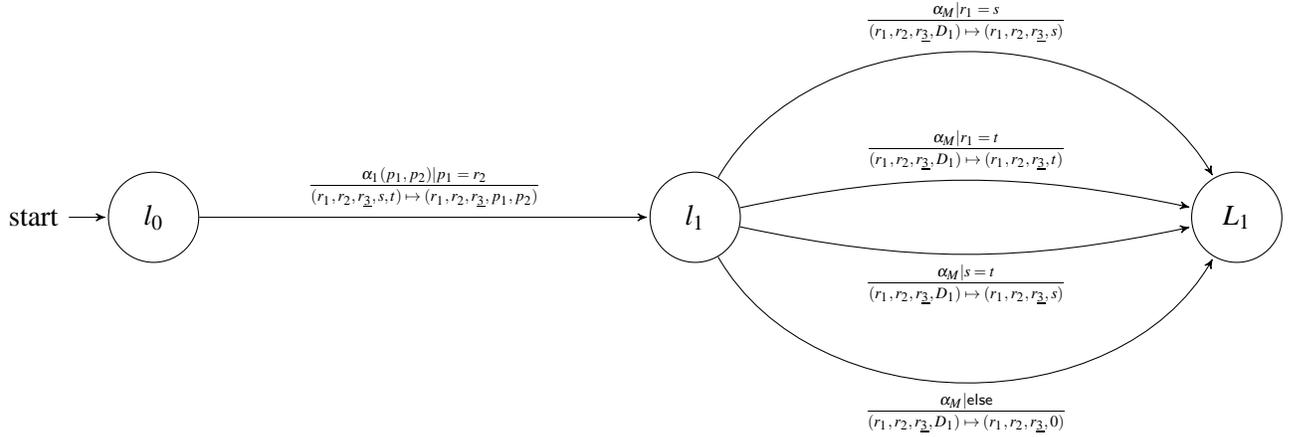

Assume the lieutenant generals have received a decision from the
commanding general initially. Since the commanding general may be
treacherous, the registers $r_1, r_2, r_3$ have arbitrary values at
location $l_0$ (Figure~\ref{figure:lg1}). 

In our scenario, the lieutenant general $1$ acts first. He receives
two messages from the other lieutenant generals in the action
$\alpha_1 (p_1, p_2)$. Since the lieutenant general $2$ is loyal, he
sends the message received from the commanding general. Thus we have
the guard $p_1 = r_2$. The message from the lieutenant general $\ug$
is arbitrary because the general is treacherous. We record the
messages from the lieutenant generals $2$ and $\ug$ in the registers
$s$ and $t$ respectively (location $l_1$). The lieutenant general $1$
makes his decision by the majority of the message from the commanding
general ($r_1$), the message from the lieutenant general $2$ ($s$),
and the message from the treacherous lieutenant general $\ug$
($t$). For instance, if the messages from the other lieutenant
generals are equal ($s = t$), the lieutenant general $1$ will have
his decision equal to $s$ through the transition $(l_1, \alpha_M, s = t, 
(r_1, r_2, r_{\ug}, D_1) \mapsto (r_1, r_2, r_{\ug}, s) ,L_1)$.

The other lieutenant generals are modeled similarly. 
Appendix~\ref{section:full-example} gives the model in register
automata for the scenario where the location $L_2$ denotes the end of
communication. Since the commanding general is not necessary loyal, we
are interested in finding initial configurations that satisfy the
$\mathit{CTL} (X, L)$ property $AF (D_1 = D_2) \equiv
\neg EG \neg (D_1 = D_2)$.

Let $\mathcal{R} = \mathsf{UniverseR} (X)$ be the set of
representative matrices. We begin with 
$U_0 = \{ \langle l, R_{[v]} \rangle | \langle l, v 
\rangle \models_{RA} \neg (D_1 = D_2) \} = L \times \{ R_{[v]} | v
(D_1) \neq v (D_2) \}$. Then 
$W_0 = \mathsf{ComputeEX} (RA, U_0) = (\{ l_0, l_1, L_1, L_3 \} \times 
\mathcal{R}) \cup \{ \langle l_2, R_{[v]} \rangle | 
(v (r_2) = v (s) \wedge v (D_1) \neq v (s)) \vee
(v (r_2) = v (t) \wedge v (D_1) \neq v (t)) \vee
(v (s) = v (t) \wedge v (D_1) \neq v (s)) \vee 
(v (r_2) \neq v (s) \wedge v (r_2) \neq v (t) \wedge v (s) \neq v (t)
\wedge v (D_1) \neq v (0)) \} \cup 
\{ \langle L_2, R_{[v]} \rangle | v (D_1) \neq v (D_2) \}$. 
Consider a configuration $\langle l_1, v_1 \rangle \in \langle l_1,
R_{[v_1]} \rangle \in W_0$. Since
the outgoing transitions at location $l_1$ do not assign values to
the register $D_2$, $D_2$ can have an arbitrary value at the location
$L_1$. Particularly, $\langle l_1, v_1 \rangle
\xrightarrow{\alpha_M} \langle l_1, v'_1 \rangle$ for some $v'_1 (D_2)
\neq v'_1 (D_1)$. We have $\langle l_1, v_1 \rangle \models_{RA} EX
\neg (D_1 = D_2)$. 
More interestingly, let us consider another configuration
$\langle l_2, v_2 \rangle \in \langle l_2, R_{[v_2]} \rangle \in W_0$
with $v_2 (s) = v_2 (t) \wedge v_2 (D_1) \neq v_2 (s)$. Since 
$v_2 (s) = v_2 (t)$, the register $D_2$ will be assigned to the
value of the register $s$ by the transition $(l_2, \alpha_M, s = t,
(r_1, r_2, r_{\ug}, D_1, D_2, D_3) \mapsto (r_1, r_2, r_{\ug},
D_1, s, D_3), L_2)$ (Figure~\ref{figure:full-example}). Particularly,
define $v'_2 (D_2) = v_2 (s)$ and $v'_2 (x) = v_2 (x)$ for $x \neq s$.
We have $\langle l_2, v_2 \rangle \xrightarrow{\alpha_M} \langle L_2,
v'_2 \rangle$, $v'_2 (D_2) = v_2 (s) \neq v_2 (D_1) = v'_2
(D_1)$, and $\langle L_2, v'_2 \rangle \models_{RA} \neg (D_1 = D_2)$.
 $\langle l_2, v_2 \rangle \models _{RA} EX \neg (D_1 = D_2)$.

We manually compute the representative configurations obtained by
$\mathsf{ComputeCTL} (RA, EG \neg (D_1 = D_2))$
(Appendix~\ref{section:computectl}). Particularly, we have 
$\{ \langle l_0, R_{[v]} \rangle | D_1 = D_2 \vee r_1 = r_2 \}
\subseteq \mathsf{ComputeCTL} (RA, AF (D_1 = D_2))$. The    
loyal lieutenant generals will agree on the same information provided
they have the same decision, or the commanding general sends them the
same message initially.  


%
%

\section{Conclusion}
\label{section:conclusion}

We develop an exact finitary representation for valuations in register
automata. Based on representative valuations, we show that the
reachability problem for register automata is decidable. We also
define $\mathit{CTL} (X, L)$ for register automata and
propose a model checking algorithm for the logic. As an illustration,
we model a scenario in the Byzantine generals problem. We discuss the
initial condition for correctness by the $\mathit{CTL} (X, L)$ model
checking algorithm in the example.

$\mathit{CTL} (X, L)$ has very primitive modal operators. We believe
that our technique applies to more expressive modal $\mu$-calculus. It
will also be interesting to investigate structured infinite
alphabets. For instance, a totally ordered infinite alphabet is useful
in the bakery algorithm. Representative valuations for such infinite
alphabets will be essential to verification as well.

\bibliographystyle{eptcs}
\bibliography{refs}

%
%

\newpage
\appendix
\section{A Scenario of the Byzantine Generals Problem}
\label{section:full-example}
\begin{figure}[h]
    \centering
\hide{
  \subfigure[Initialization]{
    \begin{tikzpicture}[->,>=stealth',shorten >=1pt,auto,
      node distance=.45\textwidth]
      \tikzstyle{every state}=     [minimum size=.075\textwidth]
    
      \node[initial, state] (1) {$l_{0}$};
      \node[state] (2) [right of=1] {$l_{c}$};

      \path[every node/.style={font=\sffamily\small}]
  
      (1) 
      edge [left] node[above] 
      {$\dfrac{\alpha_I|true}{
          \begin{matrix}
            (r_1,r_2,r_{\ug}) 
            \mapsto(r_1,r_2,r_{\ug})
          \end{matrix} 
        }$}
      (2);
    \end{tikzpicture}
  }
}


  \subfigure[Lieutenant General $1$]{
    \begin{tikzpicture}[->,>=stealth',shorten >=1pt,auto,
      node distance=.45\textwidth]
      \tikzstyle{every state}=     [minimum size=.075\textwidth]
      \node[initial, state] (2) {$l_0$};
      \node[state] (3) [right of=2] {$l_{1}$};
      \node[state] (4) [right of=3] {$L_{1}$};

      \path[every node/.style={font=\sffamily\tiny}]
  
      (2) 
      edge [left] node[above] 
      {$\dfrac{\alpha_{1}(p_1,p_2)|p_1=r_2}{
          \begin{matrix}
            (r_1,r_2,r_{\ug},s,t)
            \mapsto(r_1,r_2,r_{\ug},p_1,p_2)
          \end{matrix} 
        }$}
      (3)

      (3)
      edge [bend left=60] node[above] 
      {$\dfrac{\alpha_M|r_1=s}
        {\begin{matrix}
            (r_1, r_2, r_{\ug}, D_1) \mapsto (r_1, r_2, r_{\ug}, s)
         \end{matrix}}$}
      (4)

      (3)
      edge [bend left=12] node[above] 
      {$\dfrac{\alpha_M|r_1=t}
        {\begin{matrix}
            (r_1, r_2, r_{\ug}, D_1) \mapsto (r_1, r_2, r_{\ug}, t)
          \end{matrix}}$}
      (4)

      (3)
      edge [bend right=12] node[below] 
      {$\dfrac{\alpha_M|s = t}
        {\begin{matrix}
            (r_1, r_2, r_{\ug}, D_1) \mapsto (r_1, r_2, r_{\ug}, s)
         \end{matrix}}$}
      (4)

      (3)
      edge [bend right=60] node[below] 
      {$\dfrac{\alpha_M|\textmd{else}}
        {\begin{matrix}
            (r_1, r_2, r_{\ug}, D_1) \mapsto (r_1, r_2, r_{\ug}, 0)
         \end{matrix}}$}
      (4)
      ;
    \end{tikzpicture}
    }
  \subfigure[Lieutenant General $\ug$]{
    \begin{tikzpicture}[->,>=stealth',shorten >=1pt,auto,
      node distance=.45\textwidth]
      \tikzstyle{every state}=     [minimum size=.075\textwidth]
    
      \node[state] (4) {$L_{1}$};
      \node[state] (5) [right of=4] {$L_{\ug}$};

      \path[every node/.style={font=\sffamily\tiny}]
  
      (4) 
      edge [left] node[above] 
      {$\dfrac{\alpha_{\ug}|p_1=r_1 \wedge p_2 = r_2}{
          \begin{matrix}
            (r_1,r_2,r_{\ug},D_1) 
            \mapsto(r_1,r_2,r_{\ug},D_1)
          \end{matrix}
        }$}
      (5);
    \end{tikzpicture}
  }

  \subfigure[Lieutenant General $2$]{
    \begin{tikzpicture}[->,>=stealth',shorten >=1pt,auto,
      node distance=.45\textwidth]
      \tikzstyle{every state}=     [minimum size=.075\textwidth]
      \node[state] (5) {$L_{\ug}$};
      \node[state] (6) [right of=5] {$l_{2}$};
      \node[state] (7) [right of=6] {$L_{2}$};

      \path[every node/.style={font=\sffamily\tiny}]
  
      (5) 
      edge [left] node[above] 
      {$\dfrac{\alpha_2(p_1,p_2)|p_1=r_1}{
          \begin{matrix}
            (r_1,r_2,r_{\ug},D_1,D_{\ug},s,t)\\
            \mapsto(r_1,r_2,r_{\ug},D_1,D_{\ug},p_1,p_2)
          \end{matrix} 
        }$}
      (6)

      (6)
      edge [bend left=60] node[above] 
      {$\dfrac{\alpha_M|r_2 = s}
        {\begin{matrix}
            (r_1, r_2, r_{\ug}, D_1, D_2, D_3)
            \mapsto
            (r_1, r_2, r_{\ug}, D_1, s, D_3)
         \end{matrix}
        }$}
      (7)

      (6)
      edge [bend left=10] node[above] 
      {$\dfrac{\alpha_M|r_2 = t}
        {\begin{matrix}
            (r_1, r_2, r_{\ug}, D_1, D_2, D_3)\\
            \mapsto
            (r_1, r_2, r_{\ug}, D_1, t, D_3)
         \end{matrix}}$}
      (7)

      (6)
      edge [bend right=10] node[below] 
      {$\dfrac{\alpha_M|s = t}
        {\begin{matrix}
            (r_1, r_2, r_{\ug}, D_1, D_2, D_3)\\
            \mapsto
            (r_1, r_2, r_{\ug}, D_1, s, D_3)
         \end{matrix}}$}
      (7)

      (6)
      edge [bend right=60] node[below] 
      {$\dfrac{\alpha_M|\textmd{else}}
        {\begin{matrix}
            (r_1, r_2, r_{\ug}, D_1, D_2, D_3)
            \mapsto
            (r_1, r_2, r_{\ug}, D_1, 0, D_3)
         \end{matrix}}$}
      (7)

      (7)
      edge [loop right] node[above] 
      {$\dfrac{ - | \mathit{true}}
        {X \mapsto X}$}
      (7)
      ;
    \end{tikzpicture}
    }
  \caption{The Byzantine Generals Problem}
  \label{figure:full-example}
\end{figure}
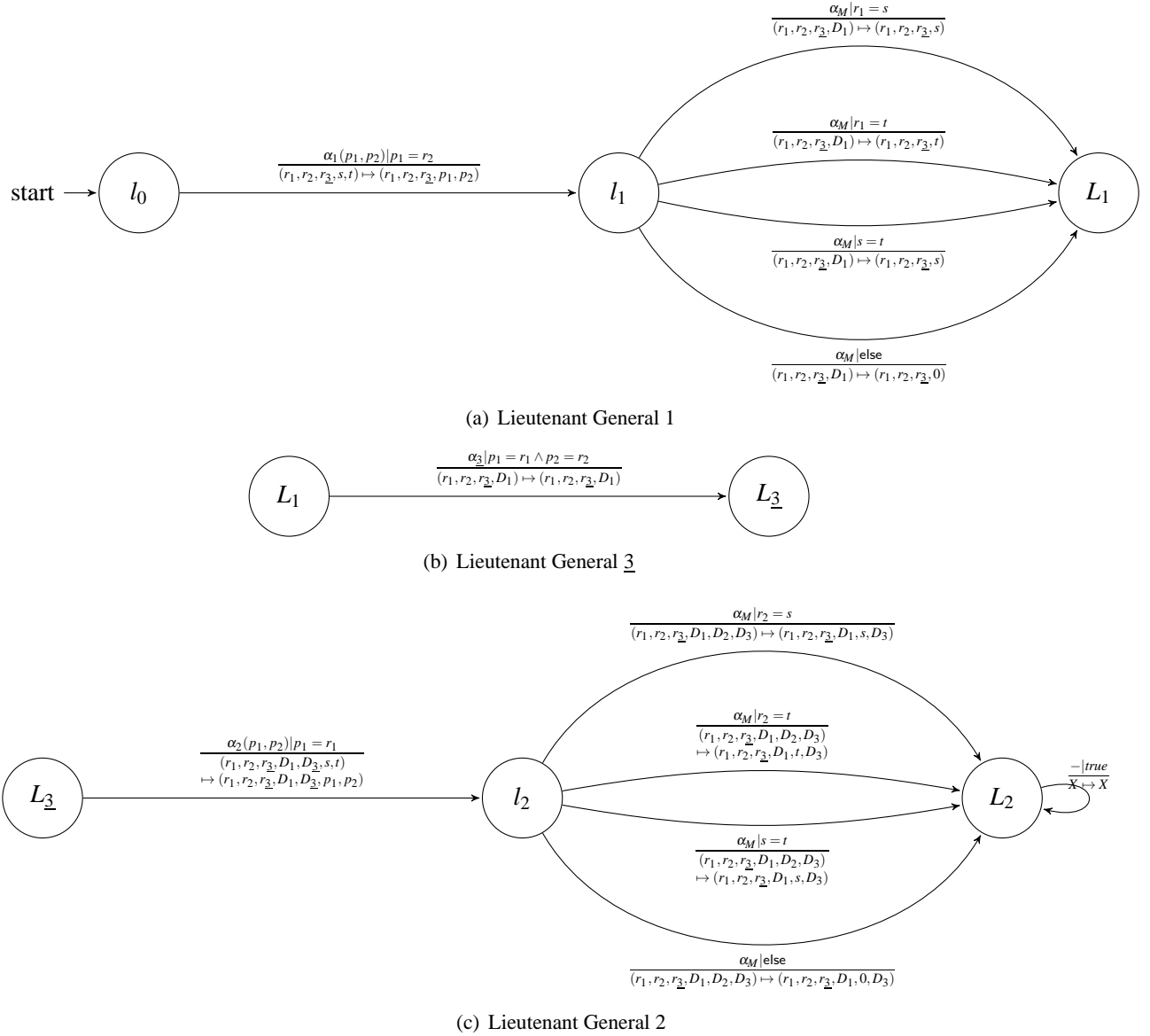
Figure~\ref{figure:full-example} shows the register automaton for the
scenario described in Section~\ref{section:example}. The transition 
$\dfrac{-| \mathit{true}}{X \mapsto X}$ at location $L_2$ denotes that
the automaton keeps the same valuation upon reading any data symbol at
location $L_2$.

\section{$\mathsf{ComputeCTL} (RA, EG \neg (D_1 = D_2))$}
\label{section:computectl}

Let $\mathcal{R} = \mathsf{UniverseR} (X)$. In the following, the set
comprehension represents the requirements of valuations. For instance,
the notation $\{ R_{[v]} | D_1 \neq D_1 \}$ denotes the set 
$\{ R_{[v]} | v (D_1) \neq v (D_2) \}$. The following table shows the
details of computation.

{
\begin{tabular}{|c|l|}
  \hline
  $U_0$ & $L \times \{ R_{[v]} | D_1 \neq D_2\}$ \\
  \hline
  $W_0$ & 
  $
  \begin{array}{lr}
    \{ l_0, l_1, L_1, L_{3} \} \times \mathcal{R} & \cup\\
    \left\{
      \begin{array}{l|l}
        \multirow{2}{*}{ $\langle l_2, R_{[v]} \rangle $ } &
        (r_2 = s \wedge D_1\neq s) \vee 
        (r_2 = t \wedge D_1 \neq t) \vee 
        (s = t \wedge D_1 \neq s) \vee \\
        & (r_2 \neq s \wedge r_2 \neq t \wedge s \neq t \wedge D_1 \neq 0)
      \end{array}
    \right\} & \cup\\
    \{ \langle L_2, R_{[v]} \rangle | D_1 \neq D_2 \}
  \end{array}
  $\\
  \hline
  $U_1$ & 
  $
  \begin{array}{lr}
    \{ l_0, l_1, L_1, L_{3} \} \times \{R_{[v]} | D_1 \neq D_2\} & \cup \\
    \left\{
      \begin{array}{l|l}
        \multirow{3}{*}{ $ \langle l_2, R_{[v]} \rangle $ } &
        (D_1 \neq D_2) \wedge \\
        &
        ((r_2 = s \wedge D_1 \neq s) \vee (r_2 = t \wedge D_1\neq t) \vee
         (s = t \wedge D_1 \neq s) \vee\\
         &
         (r_2 \neq s \wedge r_2 \neq t \wedge s \neq t \wedge D_1 \neq 0))
      \end{array}
    \right\} & \cup\\
    \{ \langle L_2, R_{[v]} \rangle | D_1 \neq D_2\}
  \end{array}
  $\\
  \hline
  $W_1$ &
  $
  \begin{array}{lr}
    \{ l_0, l_1, L_1 \} \times \mathcal{R} & \cup \\
    \{ \langle L_3, R_{[v]} \rangle |
    (D_1 \neq r_2) \vee (D_1 \neq 0 \wedge r_1\neq r_2) \} & \cup\\
    \left\{
      \begin{array}{l|l}
        \multirow{3}{*}{ $ \langle l_2, R_{[v]} \rangle $ } &
        (D_1 \neq D_2) \wedge \\
        &
        ((r_2 = s \wedge D_1 \neq s) \vee (r_2 = t \wedge D_1\neq t) \vee
         (s = t \wedge D_1 \neq s) \vee\\
         &
         (r_2 \neq s \wedge r_2 \neq t \wedge s \neq t \wedge D_1 \neq 0))
      \end{array}
    \right\} & \cup\\
    \{ \langle L_2, R_{[v]} \rangle | D_1 \neq D_2\}
  \end{array}
  $
  \\
  \hline
  $U_2$ &
  $
  \begin{array}{lr}
    \{ l_0, l_1, L_1 \} \times \{R_{[v]} | D_1 \neq D_2\} & \cup \\
    \{ \langle L_3, R_{[v]} \rangle |
    (D_1 \neq D_2) \wedge (D_1 \neq r_2) \vee
    (D_1 \neq 0 \wedge r_1 \neq r_2))\} & \cup\\
    \left\{
      \begin{array}{l|l}
        \multirow{3}{*}{ $ \langle l_2, R_{[v]} \rangle $ } &
        (D_1 \neq D_2) \wedge \\
        &
        ((r_2 = s \wedge D_1 \neq s) \vee (r_2 = t \wedge D_1\neq t) \vee
         (s = t \wedge D_1 \neq s) \vee\\
         &
         (r_2 \neq s \wedge r_2 \neq t \wedge s \neq t \wedge D_1 \neq 0))
      \end{array}
    \right\} & \cup\\
    \{ \langle L_2, R_{[v]} \rangle | D_1 \neq D_2\}
  \end{array}
  $
  \\
  \hline
  $W_2$ &
  $
  \begin{array}{lr}
    \{ l_0, l_1 \} \times \mathcal{R} & \cup\\
    \{ \langle L_1, R_{[v]} \rangle |
    (D_1 \neq r_2) \vee (D_1 \neq 0 \wedge r_1 \neq r_2) \} & \cup\\
    \{ \langle L_3, R_{[v]} \rangle |
    (D_1 \neq D_2) \wedge (D_1 \neq r_2) \vee
    (D_1 \neq 0 \wedge r_1 \neq r_2))\} & \cup\\
    \left\{
      \begin{array}{l|l}
        \multirow{3}{*}{ $ \langle l_2, R_{[v]} \rangle $ } &
        (D_1 \neq D_2) \wedge \\
        &
        ((r_2 = s \wedge D_1 \neq s) \vee (r_2 = t \wedge D_1\neq t) \vee
         (s = t \wedge D_1 \neq s) \vee\\
         &
         (r_2 \neq s \wedge r_2 \neq t \wedge s \neq t \wedge D_1 \neq 0))
      \end{array}
    \right\} & \cup\\
    \{ \langle L_2, R_{[v]} \rangle | D_1 \neq D_2\}
  \end{array}
  $
  \\
  \hline
  $U_3$ &
  $
  \begin{array}{lr}
    \{ l_0, l_1 \} \times \{ R_{[v]} | D_1 \neq D_2\} & \cup \\
    \{ \langle L_1, R_{[v]} \rangle |
    (D_1 \neq D_2) \wedge 
    ((D_1 \neq r_2) \vee (D_1 \neq r_1) \vee 
    (D_1 \neq 0 \wedge r_1 \neq r_2))\} & \cup\\
    \{ \langle L_3, R_{[v]} \rangle |
    (D_1 \neq D_2) \wedge (D_1 \neq r_2) \vee
    (D_1 \neq 0 \wedge r_1 \neq r_2))\} & \cup\\
    \left\{
      \begin{array}{l|l}
        \multirow{3}{*}{ $ \langle l_2, R_{[v]} \rangle $ } &
        (D_1 \neq D_2) \wedge \\
        &
        ((r_2 = s \wedge D_1 \neq s) \vee (r_2 = t \wedge D_1\neq t) \vee
         (s = t \wedge D_1 \neq s) \vee\\
         &
         (r_2 \neq s \wedge r_2 \neq t \wedge s \neq t \wedge D_1 \neq 0))
      \end{array}
    \right\} & \cup\\
    \{ \langle L_2, R_{[v]} \rangle | D_1 \neq D_2\}
  \end{array}
  $
  \\
  \hline
      \end{tabular}
}

{\small
\begin{tabular}{|c|l|}
\hline
  $W_3$ &
  $
  \begin{array}{lr}
    \{ l_0 \} \times \mathcal{R} & \cup \\
    \left\{ 
      \begin{array}{l|l}
        \multirow{4}{*}{ $\langle l_1, R_{[v]} \rangle$ }
        & 
        [(r_1 = s \wedge s \neq r_2) \vee (r_1 = t \wedge t \neq r_2) \vee
         (s = t \wedge s \neq r_2) \vee \\
        & \ \ \ \ 
        (r_1 \neq s \wedge r_1 \neq t \wedge s \neq t \wedge r_2 \neq 0)] 
        \vee\\
        & 
        [(r_1 = s \wedge s \neq 0 \wedge r_1 \neq r_2) \vee
         (r_1 =t \wedge t \neq 0 \wedge r_1 \neq r_2) \vee\\
        & \ \ \ \ 
         (s = t \wedge s \neq 0 \wedge r_1 \neq r_2)]
      \end{array}
    \right\}    & \cup\\
    \{ \langle L_1, R_{[v]} \rangle |
    (D_1 \neq D_2) \wedge 
    ((D_1 \neq r_2) \vee (D_1 \neq r_1) \vee 
    (D_1 \neq 0 \wedge r_1 \neq r_2))\} & \cup\\
    \{ \langle L_3, R_{[v]} \rangle |
    (D_1 \neq D_2) \wedge (D_1 \neq r_2) \vee
    (D_1 \neq 0 \wedge r_1 \neq r_2))\} & \cup\\
    \left\{
      \begin{array}{l|l}
        \multirow{3}{*}{ $ \langle l_2, R_{[v]} \rangle $ } &
        (D_1 \neq D_2) \wedge \\
        &
        ((r_2 = s \wedge D_1 \neq s) \vee (r_2 = t \wedge D_1\neq t) \vee
         (s = t \wedge D_1 \neq s) \vee\\
         &
         (r_2 \neq s \wedge r_2 \neq t \wedge s \neq t \wedge D_1 \neq 0))
      \end{array}
    \right\} & \cup\\
    \{ \langle L_2, R_{[v]} \rangle | D_1 \neq D_2\}
  \end{array}
  $
  \\
  \hline
  $U_4$ &
  $
  \begin{array}{lr}
    \{ \langle l_0, R_{[v]} \rangle | D_1 \neq D_2\} & \cup \\
    \left\{
      \begin{array}{l|l}
        \multirow{5}{*}{ $\langle l_1, R_{[v]} \rangle$ } &
        (D_1 \neq D_2) \wedge \\
        &
        ([(r_1 = s \wedge s \neq r_2) \vee 
          (r_1 = t \wedge t \neq r_2) \vee
          (s = t \wedge s \neq r_2) \vee \\
        & \ \ \ \ 
          (r_1 \neq s \wedge r_1 \neq t \wedge s \neq t \wedge r_2 \neq 0)]
        \vee\\
        &
         \ [(r_1 = s \wedge s \neq 0 \wedge r_1 \neq r_2) \vee
          (r_1 = t \wedge t \neq 0 \wedge r_1 \neq r_2) \vee \\
        & \ \ \ \ 
          (s = t \wedge s \neq 0 \wedge r_1 \neq r_2)])\\
      \end{array}
      \right\} & \cup\\
    \{ \langle L_1, R_{[v]} \rangle |
    (D_1 \neq D_2) \wedge 
    ((D_1 \neq r_2) \vee (D_1 \neq r_1) \vee 
    (D_1 \neq 0 \wedge r_1 \neq r_2))\} & \cup\\
    \{ \langle L_3, R_{[v]} \rangle |
    (D_1 \neq D_2) \wedge (D_1 \neq r_2) \vee
    (D_1 \neq 0 \wedge r_1 \neq r_2))\} & \cup\\
    \left\{
      \begin{array}{l|l}
        \multirow{3}{*}{ $ \langle l_2, R_{[v]} \rangle $ } &
        (D_1 \neq D_2) \wedge \\
        &
        ((r_2 = s \wedge D_1 \neq s) \vee (r_2 = t \wedge D_1\neq t) \vee
         (s = t \wedge D_1 \neq s) \vee\\
         &
         (r_2 \neq s \wedge r_2 \neq t \wedge s \neq t \wedge D_1 \neq 0))
      \end{array}
    \right\} & \cup\\
    \{ \langle L_2, R_{[v]} \rangle | D_1 \neq D_2\}
  \end{array}
  $
  \\
  \hline

  $W_4$ &
  $
  \begin{array}{lr}
    \{ \langle l_0, R_{[v]} \rangle | r_1 \neq r_2\} & \cup\\
    \left\{
      \begin{array}{l|l}
        \multirow{5}{*}{ $\langle l_1, R_{[v]} \rangle$ } &
        (D_1 \neq D_2) \wedge \\
        &
        ([(r_1 = s \wedge s \neq r_2) \vee 
          (r_1 = t \wedge t \neq r_2) \vee
          (s = t \wedge s \neq r_2) \vee \\
        & \ \ \ \ 
          (r_1 \neq s \wedge r_1 \neq t \wedge s \neq t \wedge r_2 \neq 0)]
        \vee\\
        &
         \ [(r_1 = s \wedge s \neq 0 \wedge r_1 \neq r_2) \vee
          (r_1 = t \wedge t \neq 0 \wedge r_1 \neq r_2) \vee \\
        & \ \ \ \ 
          (s = t \wedge s \neq 0 \wedge r_1 \neq r_2)])\\
      \end{array}
      \right\} & \cup\\
    \{ \langle L_1, R_{[v]} \rangle |
    (D_1 \neq D_2) \wedge 
    ((D_1 \neq r_2) \vee (D_1 \neq r_1) \vee 
    (D_1 \neq 0 \wedge r_1 \neq r_2))\} & \cup\\
    \{ \langle L_3, R_{[v]} \rangle |
    (D_1 \neq D_2) \wedge (D_1 \neq r_2) \vee
    (D_1 \neq 0 \wedge r_1 \neq r_2))\} & \cup\\
    \left\{
      \begin{array}{l|l}
        \multirow{3}{*}{ $ \langle l_2, R_{[v]} \rangle $ } &
        (D_1 \neq D_2) \wedge \\
        &
        ((r_2 = s \wedge D_1 \neq s) \vee (r_2 = t \wedge D_1\neq t) \vee
         (s = t \wedge D_1 \neq s) \vee\\
         &
         (r_2 \neq s \wedge r_2 \neq t \wedge s \neq t \wedge D_1 \neq 0))
      \end{array}
    \right\} & \cup\\
    \{ \langle L_2, R_{[v]} \rangle | D_1 \neq D_2\} 
  \end{array}
  $
  \\
  \hline
\end{tabular}
}

Finally, the following representative configurations satisfy $EG \neg
(D_1 = D_2)$.
\begin{equation*}
  \begin{array}{ll}
    \{ \langle l_0, R_{[v]} \rangle | D_1 \neq D_2 \wedge r_1 \neq r_2 \} & 
    \cup \\
    \left\{
      \begin{array}{l|l}
        \multirow{5}{*}{$ \langle l_1, R_{[v]} \rangle$} & 
        (D_1 \neq D_2) \wedge\\
        &
        ([(r_1 = s \wedge s \neq r_2) \vee 
          (r_1 = t \wedge t \neq r_2) \vee
          (s = t \wedge s \neq r_2) \vee\\
        & \ \ \ \ 
          (r_1 \neq s \wedge r_1 \neq t \wedge s \neq t \wedge r_2 \neq 0)]
          \vee\\
        &
        \ [(r_1 = s \wedge s \neq 0 \wedge r_1 \neq r_2) \vee
           (r_1 = t \wedge t \neq 0 \wedge r_1 \neq r_2) \vee \\
        &  \ \ \ 
           (s = t \wedge s \neq 0 \wedge r_1 \neq r_2)])
      \end{array}
    \right\} & 
    \cup \\
    \{ \langle L_1, R_{[v]} \rangle | (D_1 \neq D_2) \wedge 
                ((D_1 \neq r_2) \vee (D_1 \neq 0 \wedge r_1 \neq r_2))\} & 
    \cup\\
    \{ \langle L_3, R_{[v]} \rangle | (D_1 \neq D_2) \wedge 
                ((D_1\neq r_2) \vee (D_1 \neq 0 \wedge r_1 \neq r_2))\} &
    \cup\\
    \left\{
    \begin{array}{l|l}
      \multirow{3}{*}{ $ \langle l_2, R_{[v]} \rangle $ } & 
      (D_1 \neq D_2) \wedge\\
      &
      [(r_2 = s \wedge D_1 \neq s) \vee
       (r_2 = t \wedge D_1 \neq t) \vee
       (s = t \wedge D_1 \neq s) \vee\\
      & \ \ \ \ 
       (r_2 \neq s \wedge r_2 \neq t \wedge s \neq t \wedge D_1\neq 0)]
    \end{array}
    \right\} &
    \cup \\
    \{ \langle L_2, R_{[v]} \rangle | D_1 \neq D_2\}
  \end{array}
\end{equation*}

\end{document}